\documentclass[journal]{IEEEtran}
\usepackage{cite}
\usepackage{amsmath,amssymb,amsfonts}
\usepackage{algorithmic}
\usepackage{graphicx}
\usepackage{textcomp}
\usepackage{url}
\usepackage[breaklinks]{hyperref}
\usepackage{tabularray}
\usepackage{array}
\usepackage{makecell}
\usepackage{rotating}
\usepackage{multirow}
\usepackage[table]{xcolor}
\usepackage{booktabs}
\usepackage{comment}
\usepackage{amsthm}

\def\BibTeX{{\rm B\kern-.05em{\sc i\kern-.025em b}\kern-.08em
    T\kern-.1667em\lower.7ex\hbox{E}\kern-.125emX}}

\newif\ifshowhighlights
\showhighlightsfalse   

\usepackage{soul}

\sethlcolor{yellow} 

\ifshowhighlights
  \newcommand{\hlchange}[1]{\hl{#1}}
\else
  \newcommand{\hlchange}[1]{#1}
\fi
    
\begin{document}

\title{VerifAI: A Verifiable Open-Source Search Engine for Biomedical Question Answering}

\author{Milo\v{s}~Ko\v{s}prdi\'{c},
        Adela~Ljaji\'{c},
        Bojana~Ba\v{s}aragin,
        Darija~Medvecki,
        Lorenzo~Cassano,
        and~Nikola~Milo\v{s}evi\'{c}
\thanks{M. Ko\v{s}prdi\'{c}, A. Ljaji\'{c}, B. Ba\v{s}aragin, and D. Medvecki are with The Institute for Artificial Intelligence Research and Development of Serbia, Fru\v{s}kogorska 1, 21000 Novi Sad, Serbia (e-mail: \{milos.kosprdic, adela.ljajic, bojana.basaragin, darija.medvecki\}@ivi.ac.rs).}
\thanks{L. Cassano is with Bayer A.G., Muellerstrasse 178, 13353 Berlin, Germany (e-mail: lorenzo.cassano@bayer.com).}
\thanks{N. Milo\v{s}evi\'{c} is with The Institute for Artificial Intelligence Research and Development of Serbia, Novi Sad, Serbia, and also with Bayer A.G., Berlin, Germany (e-mail: nikola.milosevic@ivi.ac.rs, nikola.milosevic@bayer.com).}
\thanks{Manuscript received January 16, 2026.}}

\markboth{IEEE ACCESS}%
{Ko\v{s}prdi\'{c} \textit{et al.}: VerifAI: A Verifiable Open-Source Search Engine for Biomedical Question Answering}

\maketitle

\begin{abstract}
We introduce VerifAI, an open-source expert system for biomedical question answering that integrates retrieval-augmented generation (RAG) with a novel post-hoc claim verification mechanism. Unlike standard RAG systems, VerifAI ensures factual consistency by decomposing generated answers into atomic claims and validating them against retrieved evidence using a fine-tuned natural language inference (NLI) engine. The system comprises three modular components: (1) a hybrid Information Retrieval (IR) module optimized for biomedical queries (MAP@10 of 42.7\%), (2) a citation-aware Generative Component fine-tuned on a custom dataset to produce referenced answers, and (3) a Verification Component that detects hallucinations with state-of-the-art accuracy, outperforming GPT-4 on the HealthVer benchmark. Evaluations demonstrate that VerifAI significantly reduces hallucinated citations compared to zero-shot baselines and provides a transparent, verifiable lineage for every claim. The full pipeline, including code, models, and datasets, is open-sourced to facilitate reliable AI deployment in high-stakes domains.
\end{abstract}


\maketitle

\section{Introduction}
\label{sec:introduction}
Text-based generative AI has made a significant impact across various aspects of our daily lives, from content creation to customer service and beyond \cite{achiam2023gpt,katz2024gpt,bubeck2023sparks,nori2023capabilities}. Its ability to generate human-like text has revolutionized the way we access and interact with information. Within the scientific community, these models offer significant promise in speeding up research workflows, streamlining information retrieval, and improving the creation of sophisticated scientific materials. However, the widespread adoption of generative AI, particularly in domains that require verifiable information, is hindered by the problem of hallucinations of large language models (LLMs) \cite{ji2023survey}. Hallucinations refer to instances of generated text that are incorrect or nonsensical, despite being presented eloquently and convincingly. This issue is of special concern in life sciences, where the factuality and truthfulness of answers are vital \cite{tsatsaronis2015overview,wadden_fact_2020}.

The prevalence of hallucinations is directly correlated with the quantity and quality of data on which the models are trained \cite{huang2025survey}. Despite the vast amounts of training data, even the largest LLMs are still undersourced, especially in specialized domains \cite{peskoff-stewart-2023-credible}. The risk of misinformation creates a fundamental trust gap, preventing users from fully adopting generative language models. To safeguard the integrity of scientific knowledge while maximizing the advantages these models offer, it is essential to tackle this issue.

We propose VerifAI, an expert system based on generative search and natural language inference, designed to generate verifiable answers to biomedical questions. VerifAI builds upon the principles of adaptive retrieval strategies and context relevance assessment by integrating citation generation and claim verification. VerifAI generates answers supported by established sources, namely PubMed abstracts, while simultaneously offering citations for the claims and verifying if the generated claims correspond to the original content. This engine not only provides users with a clearer understanding of the origins of the information, which is freely available online and credible, but also adds a critical layer of verification by checking whether each generated claim is rooted in the provided reference. By enabling users to receive a verified answer to their inquiries, VerifAI aims to mitigate the spread of disinformation and misinformation on the web, ultimately enhancing the integrity of the information accessed by users and increasing public trust in generative AI.

\hlchange{This work introduces three key methodological contributions: (1) A fine-tuning strategy that enables Small Language Models (SLMs) to achieve citation fidelity comparable to frontier LLMs, challenging the necessity of massive models for verifiable QA; (2) An empirical demonstration that specialized natural language inference (NLI) discriminators significantly outperform general-purpose generative verification (including GPT-4) on biomedical benchmarks; and (3) The release of the first open-source, end-to-end verifiable QA pipeline that integrates hybrid retrieval, citation-aware generation, and post-hoc entailment verification in a modular framework.}

The remainder of this paper is organized as follows. We begin by providing background on the components and theoretical underpinnings of our approach (Section~\ref{sec:background}). The architecture of the VerifAI pipeline and its key components are then described in Section~\ref{sec:methods}. Section~\ref{sec:results} presents the results of both standalone and end-to-end evaluations. The paper concludes with a discussion of key findings (Section \ref{sec:discussion}), comparisons with other systems (Section \ref{sec:comparison}), extensibility and adaptation to other domains (Section \ref{sec:generalization}) and final remarks (Section \ref{sec:conclusion}), while also addressing the limitations of our approach (Section \ref{sec:limitations}). In the spirit of open science, all our code, models, and datasets have been made publicly available, with details provided in Section \ref{sec:availability}.

\section{Background and Related Work}
\label{sec:background}

Large language models (LLMs) have demonstrated remarkable capabilities across a range of natural language tasks, including biomedical question answering (QA). However, their responses often suffer from hallucinations-plausible-sounding but factually incorrect or unsupported statements \cite{ji2023survey, barnett2024seven}. In biomedical and clinical contexts, such errors can have serious consequences, making factual consistency and verifiability essential.

Despite domain-specific pretraining efforts such as BioGPT \cite{luo2022biogpt}, PubMedGPT, PMC-LLama \cite{wu2024pmc} or Qibo \cite{JIA2025127672}, hallucinations persist due to both the complexity of biomedical knowledge and the limitations of training data coverage. Retrieval-augmented methods and verification mechanisms have been proposed to address these challenges, but most existing biomedical QA systems still fall short in ensuring both answer quality and verifiability.

Retrieval-augmented generation (RAG) is a widely adopted framework for improving factual grounding in LLMs \cite{lewis2020retrieval}. By retrieving relevant external documents to condition generation, RAG reduces reliance on the model's parametric knowledge and increases transparency. Biomedical implementations of this strategy include RAG-enabled versions of PubMedGPT and tools like Elicit, which retrieve from indexed biomedical sources to improve relevance.

However, while RAG can improve the factuality of answers, it does not guarantee alignment between retrieved evidence and generated responses. Studies have shown that even with access to accurate documents, LLMs may still introduce unsupported or misleading claims \cite{barnett2024seven}. Moreover, citation fidelity remains problematic: Gao et al. \cite{gao2023enabling} and Liu et al. \cite{liu2023evaluating} found that only around 50\% of generated sentences in biomedical RAG systems are properly cited, and only 75\% of those citations actually support the associated claims.

To address the limitations of RAG, researchers have introduced verification modules that assess whether generated claims are supported by retrieved evidence. These modules typically rely on  NLI --- the task of determining whether a hypothesis (e.g., a model-generated statement) is entailed, contradicted, or unrelated to a given premise (e.g., the retrieved source), within a closed, evidence-conditioned context. In the biomedical domain, datasets such as SciFact \cite{wadden_fact_2020} and BioNLI \cite{bastan2022bionli} have been used to train and evaluate such systems.

\hlchange{It is important to note that the type of verification considered in this work focuses on entailment-based claim verification, where each generated statement (hypothesis) is evaluated for logical support against explicitly retrieved evidence (premise). This differs from broader fact-checking approaches, which may consider open-domain truthfulness or rely on external knowledge without conditioning strictly on retrieved documents.}

Verification mechanisms can be integrated at various stages of the pipeline:
\begin{itemize}
\item \textbf{Pre-generation}, e.g., filtering context with the FilCo method \cite{Wang2023Learning};
\item \textbf{During generation}, e.g., explicit working memory models \cite{chen2025improvingfactualityexplicitworking}, or reflection-based prompting \cite{Asai2023Self};
\item \textbf{Post-generation}, e.g., entailment-based claim verification like FACTOID \cite{Rawte2024FACTOID}.
\end{itemize}

Some of these methods report substantial improvements: FACTOID boosts verification accuracy by 40\% over standard entailment models, and VERA achieves citation-quality gains of 9–22\% over reranking baselines \cite{Birur2024VERA, Huo2023Retrieving}.

More complex architectures include:
\begin{itemize}
\item \textbf{Hierarchical verification}, decomposing questions into sub-claims \cite{Fang2024HGOT};
\item \textbf{Self-reflection}, which limits unsupported answers to as low as 2\% \cite{Asai2023Self};
\item \textbf{Adaptive retrieval}, which adjusts the retrieval process based on reasoning complexity \cite{Xu2024CRP, Yan2024Corrective}.
\end{itemize}

Tang et al. \cite{tang2025chatsos}  propose ChatSOS, a vector database augmented generative system for safety engineering queries. By constructing a vector database from historical accident reports, their approach significantly improves the reliability, accuracy, and adaptability of LLM-generated answers in a professional domain. This highlights the value of retrieval-enhanced QA --- complementary to VerifAI’s RAG approach. Although effective, most of these techniques have been evaluated in isolation or general scientific QA settings. To our knowledge, no prior biomedical QA pipeline has jointly implemented retrieval, generation, citation production, and post-hoc claim verification using entailment reasoning in a single system.

\hlchange{Recent comprehensive surveys have provided systematic overviews of RAG applications in biomedicine, complementing the foundational RAG work discussed earlier. An extensive survey of RAG technologies, datasets, and clinical applications in the biomedical domain analyzes current state-of-the-art approaches and identifies key challenges specific to medical RAG systems }  \cite{he2025retrieval}. \hlchange{A systematic review of retrieval-augmented generation for large language models in healthcare further provides evidence-based guidelines for implementing RAG systems in clinical settings} \cite{amugongo2025retrieval}. \hlchange{In addition, systematic guidelines for improving biomedical large language model applications through RAG have been proposed, supported by meta-analysis results that inform best practices for clinical development }
\cite{10.1093/jamia/ocaf008}.

\hlchange{Recent benchmarking efforts have significantly advanced the evaluation landscape for biomedical RAG. Comprehensive benchmarks specifically designed for retrieval-augmented generation in medicine have introduced standardized evaluation protocols that complement the BioASQ evaluation approach used in this work} \cite{xiong2024benchmarkingrag}. \hlchange{While not specifically biomedical, RAGTruth -- a hallucination corpus, provides valuable insights into hallucination detection methodologies that could inform verification approaches, such as the one presented in VerifAI} \cite{niu-etal-2024-ragtruth}.

\hlchange{The importance of hallucination mitigation in healthcare-oriented generative systems has also been demonstrated in applied clinical contexts. Approaches for reducing hallucinations in generative AI chatbots for cancer-related information have been developed and evaluated, showing measurable improvements in response reliability in patient-facing settings} \cite{nishisako2025reducing}. \hlchange{A complementary analysis of techniques and challenges in mitigating hallucinations in medical language models further outlines unresolved limitations and evaluation gaps, providing insights that align with the verification strategy adopted in VerifAI} \cite{pham2024towards}.

\hlchange{Recent developments in biomedical claim verification have introduced approaches that extend beyond the SciFact and HealthVer datasets used in earlier work. An evidence-based medical fact-checking system has been proposed to verify health claims using retrieved scientific literature}
\cite{vladika2024healthfc}, \hlchange{while retrieval-augmented scientific claim verification methods have been developed that directly align with the verification component employed in our VerifAI} \cite{liu2024retrieval}.


\hlchange{Several recent frameworks have further advanced biomedical RAG systems by introducing post-retrieval auditing and multi-evidence reasoning mechanisms. A post-retrieval auditing system for scientific study summaries, VERIRAG, incorporates methodological assessment before claim verification} \cite{mohole2025veriragpostretrievalauditingscientific}, \hlchange{while multi-evidence guided answer refinement, MEGA-RAG, has been proposed to mitigate hallucinations in public health applications} \cite{xu2025mega}. \hlchange{Additional work has demonstrated practical deployment of retrieval-augmented generation chatbots for orthopedic and trauma surgery patient education, highlighting real-world considerations for biomedical QA} \cite{baur2025development}.

\hlchange{These developments underscore the growing recognition of the importance of verification and hallucination mitigation in biomedical RAG systems. While many existing approaches focus on retrieval optimization or isolated fact-checking tasks, VerifAI distinguishes itself by integrating hybrid retrieval, citation-aware generation, and post-hoc entailment-based claim verification within a single, end-to-end biomedical question answering pipeline.}

The combination of factual grounding via retrieval and rigorous claim verification represents a promising yet underexplored direction in biomedical QA. VerifAI builds on this motivation by integrating a modular RAG pipeline with an entailment-based verification engine that explicitly checks whether each generated claim is supported by the retrieved evidence. In doing so, it addresses a critical gap between citation generation and factual consistency, aiming to reduce hallucinations and improve transparency in biomedical generative QA.

\begin{figure*}[h!]
     \centering
     \includegraphics[width=1.0\linewidth]{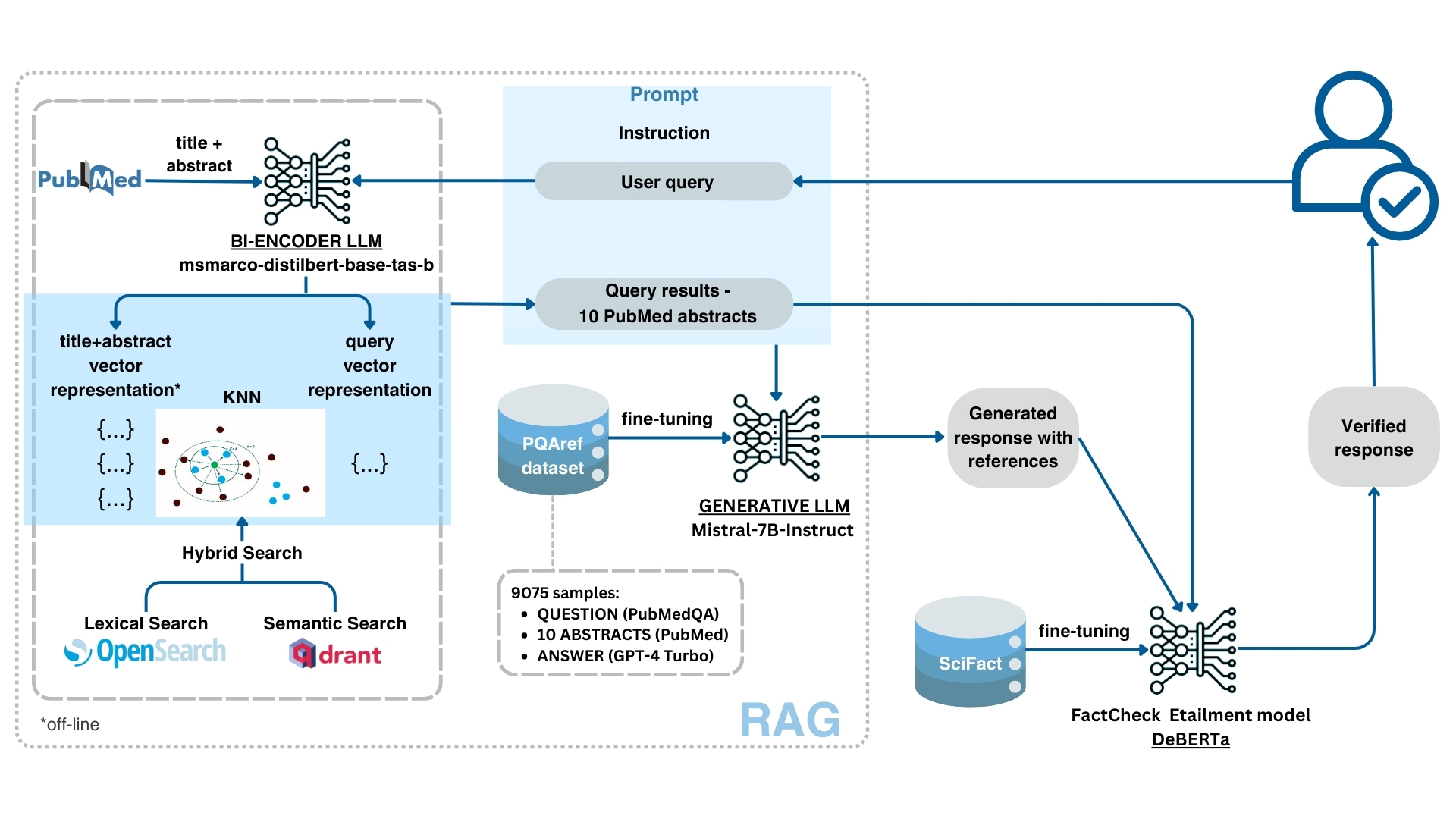}
     \caption{The architecture of VerifAI.}
     \label{Fig:system}
\end{figure*}

\section{Materials and Methods}
\label{sec:methods}

The VerifAI contains three components. The first two components form VerifAI's RAG system: the Information Retrieval Component, responsible for retrieving the most relevant scientific paper abstracts from PubMed in response to user queries, and the Generative Component, designed to produce concise answers based on the retrieved abstracts, providing references for each generated claim. The third component is a Verification Engine that cross-checks the generated answers with the referenced abstracts to ensure accuracy and identify potential hallucinations. Each of the components is presented in detail in the following subsections.

\subsection{The Information Retrieval Component}
\label{subsec:IR}

Our Information Retrieval Component is designed to process large-scale biomedical literature efficiently. Given the vast number of scientific publications, we selected PubMed \cite{pubmed} as our primary database due to its comprehensive coverage of biomedical research. We included all available articles up to February 2024 (approximately 36.8 million) and, after filtering out empty abstract entries, retained around 25.5 million abstracts for indexing.

To construct a high-quality retrieval dataset, we merged article titles and abstracts into a single text while incorporating metadata such as author names, publication dates, and journal names to facilitate filtering. This approach ensures that our retrieval system operates on a robust dataset, maximizing relevance and accuracy.

We implemented both lexical and semantic indexing, generating the indices to support hybrid retrieval. User queries undergo pre-processing before retrieval, with different techniques applied depending on the search method:
\begin{itemize}
    \item Lexical retrieval: Stop-word removal and text normalization improve the matching accuracy.
    \item Semantic retrieval: Queries are converted into dense embeddings, allowing for similarity-based document retrieval by measuring their proximity within a high-dimensional vector space.
\end{itemize}

The hybrid search mechanism integrates both approaches, ranking documents based on relevance. This combined method leverages the strengths of semantic understanding and precise keyword matching, resulting in more comprehensive and contextually relevant search results across diverse query types \cite{kuzi2020leveraging}. The retrieved documents then serve as input to the Generative Component, enhancing its ability to provide precise and contextually relevant responses.

\subsubsection{Lexical Retrieval}

For lexical indexing, we used OpenSearch\footnote{\url{https://opensearch.org/}}, an open-source distributed search and analytics engine, implementing the Best Matching 25 (BM25) ranking function \cite{robertson2009probabilistic}. BM25 is widely used for document ranking in information retrieval, as it effectively scores documents based on term frequency and inverse document frequency (TF-IDF).

Our preprocessing pipeline for lexical retrieval includes:
\begin{itemize}
    \item stopword removal --- to reduce noise;
    \item lowercasing and text normalization --- to standardize queries and indexed documents;
    \item tokenization --- to enhance search accuracy.
\end{itemize}
This method ensures efficient retrieval of keyword-based matches while maintaining computational efficiency.

\subsubsection{Semantic Retrieval}

For semantic retrieval, we utilize Qdrant\footnote{\url{https://qdrant.tech/}}, a vector database selected for its memory-mapped storage capabilities, which significantly reduce RAM usage. To optimize retrieval speed, we apply 8-bit quantization during retrieval while retaining full-precision embeddings for rescoring when necessary.

Vector similarity searches are performed using the Hierarchical Navigable Small World (HNSW) algorithm \cite{malkov2018efficient} for Approximate Nearest Neighbors (ANN) \cite{indyk1998approximate}, leveraging dot product metrics. This approach enables rapid and precise similarity searches across high-dimensional embeddings \cite{malkov2018efficient}.

Vector embeddings are generated using a bi-encoder sentence transformer model, which at the time of indexing demonstrated state-of-the-art performance in the Passage Retrieval Task\footnote{\url{https://www.sbert.net/docs/pretrained-models/msmarco-v3.html}}. Specifically, we used a transformer model pre-trained on the MS MARCO dataset \cite{Hofstaetter2021_tasb_dense_retrieval}, designed for asymmetric semantic search\footnote{\url{https://huggingface.co/sebastian-hofstaetter/distilbert-dot-tas_b-b256-msmarco}}, as shown in Figure \ref{Fig:system}.

Since the average length of concatenated titles and abstracts was 650 tokens, exceeding the model's 512 tokens limit, we implemented a sentence-based segmentation strategy. Long abstracts were split at sentence boundaries closest to the 512th token, ensuring that each segment retained semantic coherence before indexing.

\subsubsection{Hybrid Retrieval}

To improve retrieval precision and recall, we implemented a hybrid search strategy that combines lexical and semantic retrieval. To ensure comparability between the scores from the search methods, a normalization process is applied. This is achieved by performing a search, identifying the top score from each respective search engine (OpenSearch and Qdrant), and then dividing other results from the given search method by this top score. This way, scores of both retrieval methods are normalized to a scale of [0,1], allowing a standardized comparison.

The final retrieval score is computed using a weighted combination of the normalized scores, allowing us to balance lexical precision with semantic flexibility. 
\begin{equation}
    \text{hybrid\_score} = \alpha \cdot \text{lexical\_score} + \beta \cdot \text{semantic\_score}
    \label{eq:hybrid_score}
\end{equation}

Where $\alpha$ and $\beta$ are manually adjustable weights for lexical and semantic scores, and:

\begin{equation}
    \text{lexical\_score} = \frac{\text{lexical\_score\_for\_document}}{\text{max\_lexical\_score\_for\_given\_search}} \label{eq:lexical_score}
\end{equation}

\begin{equation}
    \text{semantic\_score} = \frac{\text{semantic\_score\_for\_document}}{\text{max\_semantic\_score\_for\_given\_search}} \label{eq:semantic_score}
\end{equation}

This approach ensures that:
\begin{itemize}
    \item The exact keyword matches are prioritized when available.
    \item Conceptually similar text segments can still be retrieved even if they lack direct keyword overlap.
\end{itemize}

To optimize performance, we experimented with different weight distributions, ensuring that their sum remained \textbf{1}. This analysis helped determine the optimal balance between lexical and semantic retrieval, maximizing overall retrieval effectiveness.

\hlchange{The evaluation, presented in detail in Section} \ref{subsec:IR_component_stamdalone}\hlchange{, was performed by comparing the retrieved PubMed IDs (PMIDs) against the gold-standard PMIDs in the BioASQ dataset }\cite{krithara2023bioasq}. \hlchange{We used the training dataset for the BioASQ Task 12b}\footnote{\url{https://participants-area.bioasq.org/accounts/login/?next=/Tasks/12b/trainingDataset/}}\hlchange{, which tackles biomedical semantic QA, and involves IR and summarization. The dataset contains 5,049 expert-supplied questions in English, along with links to manually selected most relevant articles retrieved from PubMed, key snippets from those articles, and gold-standard expert-written ideal answers based on the snippets.}

\subsection{The Generative Component}
\label{subsec:GEN}

The generative component is responsible for producing a concise, citation-backed answer to the user query. It receives as input, besides the user query, the 10 most relevant abstracts retrieved by the IR module. During answer generation, the model dynamically reassesses the input documents and selectively incorporates only those it deems useful for answering the question. Each factual claim in the response is followed by a citation in the form of a PubMed ID.

As the generative model we used a custom, fine-tuned Mistral-7B-Instruct-v0.2\footnote{\url{https://huggingface.co/mistralai/Mistral-7B-Instruct-v0.2}}, a 32k context window, instruct-tuned variant of Mistral-7B-v0.2. 
Despite its relatively small size, Mistral-7B-v0.2 outperforms larger models such as LLaMA 2 13B and LLaMA 1 34B, particularly on reasoning, mathematics, and code generation tasks \cite{jiang2023mistral}, making it a suitable candidate given our computational constraints.

To improve the performance on the task of referenced question-answering (QA), we fine-tuned Mistral-7B-Instruct-v0.2 using our custom-made PQAref dataset\footnote{\url{https://huggingface.co/datasets/BojanaBas/PQAref}} \cite{basaragin-etal-2024-how} consisting of 9,075 questions randomly selected from PubMedQA \cite{jin2019pubmedqa}, each paired with 10 relevant abstracts retrieved from PubMed using a combination of free text and entity-based search, and a GPT-4 Turbo generated answer. The answer summarizes the most relevant abstracts in up to $\sim$1200 tokens and cites them accordingly after each generated claim. An example of an answer follows:

\noindent
\qquad \\\fbox{
    \parbox{0.95\linewidth}{
Yes, the choice of local anesthetic can have an impact on ambulatory surgery perioperative costs. A study comparing the use of chloroprocaine 3\% to mepivacaine 1.5\% for popliteal block in outpatient foot surgery found that chloroprocaine had significantly shorter onset and block duration times, which translated to faster discharge times and reduced total perioperative costs. Despite chloroprocaine being more expensive, its use could reduce length of stay in outpatient patients and potentially increase the revenues of an ambulatory surgical center by allowing for additional cases to be scheduled (PUBMED:27290959).
Additionally, the use of local anesthesia with sedation was found to be the most cost-effective technique for anorectal surgery in the ambulatory setting, with lower intraoperative and recovery costs, shorter times to oral intake and home-readiness, and high patient satisfaction (PUBMED:11046210).
Furthermore, the implementation of a hysterectomy Enhanced Recovery After Surgery (ERAS) protocol, which includes the choice of anesthetic medications, resulted in significantly lower total medication costs and a shorter length of stay compared to the pre-ERAS group (PUBMED:33052053).
These findings suggest that the selection of anesthetic agents and techniques can influence not only clinical outcomes but also the economic aspects of ambulatory surgical care.
       }
    }
\\


The PQAref split was 80:10:10, with most inputs in the size range of 4000 to 6000 tokens in all three splits. To be able to fine-tune the model with the available resources of a single DGX NVIDIA A100-40GB GPU, we used QLoRA \cite{dettmers2023qlora} with 4-bit quantization, and tuning for standard loss, LoRA rank of 64, alpha of 16, and a dropout of 0.1, resulting in 27.2 M trainable parameters. The model was fine-tuned over 2 epochs with a learning rate of 2e-4, a batch size of 1, gradient accumulation steps of 8, and 3 warmup steps, utilizing fp16 16-bit mixed precision training. The full training run lasted approximately 32 hours, with the best checkpoint selected via evaluation loss. \hlchange{The evaluation results on the PQAref test set are presented in Section }\ref{subsec:generative_component_standalone}.

We also fine-tuned the earlier \textit{Mistral-7B-Instruct-v0.1} and compared its performance to version 0.2 using the same test set. Although both models achieved comparable results on some metrics, version 0.2 produced significantly fewer hallucinated PMIDs and referenced the most relevant abstract 2.3\% better. Additionally, its answers usually included more information. Version 0.1 had better recall in a small-scale manual analysis, but these gains did not outweigh its citation-related shortcomings. Therefore, Mistral-7B-Instruct-v0.2 was selected as the final generator for VerifAI.
\hlchange{The details of this specific evaluation are provided} in the paper \cite{basaragin-etal-2024-how}.

\subsection{The Verification Component}
\label{subsec:VER}

Similar to structured prompt-based generation \cite{wei2024longform}, our generative component produces long-form answers composed of multiple factual claims, each accompanied by citations to relevant PubMed abstracts from our IR component. This structure promotes transparency and enhances the credibility of the responses.

To ensure that these claims are genuinely supported by the cited literature, we introduce a Verification Engine that performs claim-level factuality assessment. Specifically, for each generated claim \(c\), the engine retrieves the referenced evidence \(e\) from the IR module using the cited PubMed ID and classifies the relationship between the two as one of: \textit{support}, \textit{contradict}, or \textit{no evidence}. This task is formulated as a three-class sequence classification problem based on textual entailment. Given, that the transformer models have outperformed other types of models for similar tasks, including detection of biases \cite{rodrigo2024systematic}, we fine-tuned several transformer models for this task.

The verification module plays a critical role in detecting hallucinations and validating whether cited documents truly support the model’s outputs-thereby ensuring factual consistency between the retrieved evidence and the generated response.

\hlchange{It is important to note that in our framework, a "claim" is defined as any distinct factual assertion extracted from the generated answer. This encompasses both high-level semantic conclusions (e.g., "The drug is effective") and granular factual details (e.g., specific dates, dosages, numerical values, or entity names). Consequently, the Verification Engine is tasked with identifying hallucinations at both the semantic and granular levels.}

\subsubsection{Dataset Construction and Preprocessing}

We used the SciFact dataset\footnote{\url{https://huggingface.co/datasets/allenai/scifact}} \cite{wadden_fact_2020} as the primary source of labeled claim-evidence pairs in the biomedical and scientific domains. \hlchange{SciFact consists of 1,409 expert-written claims derived from citation sentences in scientific articles, paired with the corresponding cited article abstracts. Each claim-abstract pair is expert-annotated using one of three labels: “Supports”, where a claim is supported by the abstract, “Refutes”, signifying that the abstract refutes the claim, or “Noinfo”, where no info on the claim is provided in the abstract.} However, SciFact is designed for sentence-level fact verification and does not provide an official test split. Each abstract in the dataset can contain multiple labeled sentences, which are treated independently. This makes it poorly suited for our setting, where the generative model cites entire abstracts (via PubMed IDs) and the verification component needs to assess whether the full cited document supports a claim.

To adapt SciFact to this setting, we concatenated the abstract title and body into a single document per example and grouped all sentence-level labels referring to the same abstract. Redundant entries were removed to avoid repetition of identical claim-evidence-label triples. We also applied light cleaning to the texts, including removing line breaks, excessive whitespace, and residual markup artifacts, to ensure uniform formatting. The resulting dataset contains 1,213 unique claim–evidence pairs.

Furthermore, since SciFact offers only training and development sets, we merged the two and randomly split the full set into training (80\%), validation (10\%), and test (10\%) subsets for fine-tuning and evaluation. The final dataset is publicly available via HuggingFace\footnote{\url{https://huggingface.co/datasets/MilosKosRad/SciFact_VerifAI}}, and a detailed account of this transformation process and its motivations is provided in \cite{kosprdic-etal2024verifaikmis}.

\subsubsection{Model Selection and Training}

We fine-tuned four transformer-based models for this task: RoBERTa-large \cite{liu2019roberta}, XLM-RoBERTa Large \cite{conneau2019unsupervised}, DeBERTa Large \cite{he2020deberta}, and DeBERTa-SQuAD (a DeBERTa model fine-tuned on the SQuAD dataset). These were selected based on their strong performance in prior claim verification benchmarks: DeBERTa for its effectiveness on entailment tasks \cite{tan2023multi2claim}, RoBERTa for its modified attention mechanisms and its strongest performance on the label prediction task \cite{wadden_fact_2020}, XLM-RoBERTa for multilingual adaptability and even better performance on English data compared to RoBERTa \cite{conneau2019unsupervised}, and DeBERTa-SQuAD for transferability from QA tasks.

Input formatting was tailored to model-specific tokenization:
\begin{itemize}
\item \texttt{BERT-style models:} \texttt{[CLS] claim [SEP] evidence [SEP]}
\item \texttt{RoBERTa-style models:} \texttt{<s> claim </s></s> evidence </s>}
\end{itemize}

Training was performed using the ADAM optimizer with a learning rate of 1e-5 and weight decay of 0.01. Each model was trained for up to 15 epochs with early stopping based on F1-score.

To select the final model, we evaluated all four on three test settings:
(1) the transformed SciFact test split,
(2) the HealthVer test set for domain robustness, and
(3) GPT-4–based judgments on real VerifAI generations.

\hlchange{HealthVer}\footnote{\url{https://github.com/sarrouti/HealthVer}} \cite{sarrouti_evidence-based_2021} \hlchange{is a dataset created for evidence-based fact-checking of health-related claims, consisting of 14,330 evidence-claim pairs. The claims represent the retrieved real-world claims from snippets returned by a search engine for COVID-19-related questions. The evidence statements are manually extracted from automatically retrieved relevant scientific papers. The claim-evidence relations were manually annotated as “Support” or “Refute”, depending on whether the claim is supported or refuted by the evidence, or “Neutral” if there is insufficient information to decide.}

The best-performing model across these benchmarks was deployed as the core of VerifAI’s Verification Engine. \hlchange{The user interface employs a color-coded text system to visually communicate the verification status of each claim. Supported sentences are rendered in a dark green font, while partially supported claims appear in dark yellow/orange. Contradictions are distinctly highlighted in red text, and sentences lacking references are displayed in dark gray.}

Full evaluation results are presented in Section~\ref{subsec:aloneVEresults}.

\section{Results}
\label{sec:results}

To thoroughly evaluate our proposed system, we first conducted a series of standalone evaluations for each of the three main components: Information Retrieval (IR), Generative Component (GC), and Verification Component (VC). We evaluated each component on task-specific benchmarks to identify the most effective model configurations and training strategies within its module. Specifically, we used the BioASQ dataset for the IR component, the PQAref test set for the generative module, and both the transformed SciFact and HealthVer datasets for the verification module.

Based on these evaluations, we selected the best-performing architecture and training setup for each component. We then performed a comprehensive end-to-end evaluation to assess how the system functions as a whole. This evaluation was conducted using a controlled subset of BioASQ questions with exactly 10 known relevant abstracts, allowing us to measure how well the integrated pipeline performs on a real-world referenced question answering task, including document retrieval, answer generation, and claim-level verification.

\subsection{Information Retrieval Component Evaluation}
\label{subsec:IR_component_stamdalone}

We evaluated the information retrieval (IR) component by calculating performance metrics based on the retrieval of up to 10 relevant documents for each query. The retrieval limit was set to 10 documents per query to align with the requirements of our generative component, which is designed to process exactly 10 articles per query. For evaluation, we compared the PubMed IDs (PMIDs) retrieved by our system with the gold-standard PMIDs provided in the BioASQ dataset. The comparison was quantified using precision and mean average precision (MAP) metrics, labeled as P@(Rel-Norm) and MAP@(Rel-Norm), respectively. These metrics measure the proportion of relevant PMIDs retrieved by our system relative to the total number of PMIDs retrieved. Both metrics were computed based on the set of up to 10 relevant documents available for each query. Because the number of relevant documents varies across queries in BioASQ, we normalized the metrics by dividing by the maximum number of relevant documents per query. This normalization reflects the practical scenario in which queries have differing numbers of relevant documents, providing a more accurate and standardized assessment of the IR component’s performance, particularly in the context of imbalanced relevance counts.

\begin{table}[h]
    \centering
    \caption{Our IR and PubMed search engine performance evaluation on the BioASQ dataset.
    }
    \label{tab_IR_performance}
    \resizebox{\columnwidth}{!}{%
    \begin{tabular}{lcccc}
\hline
         & P@(Rel-Norm) & MAP@(Rel-Norm) & time [ms] \\
\hline
        1. Semantic without rescore & 14.00\% & 25.70\%  & 246  \\
        2. Semantic with rescore & 14.40 \% & 26.10\% & 254\\
        3. Hybrid with rescore (lex. 0.1 sem. 0.9) & 24.70\% & 32.50\% & 442\\
        4. Hybrid with rescore (lex. 0.2 sem. 0.8) & 24.70\% & 32.50\% & 442\\
        5. Hybrid with rescore (lex. 0.3 sem. 0.7) & 24.70\% & 32.50\% & 442\\
        6. Hybrid with rescore (lex. 0.4 sem. 0.6) & 24.70\% & 32.60\% &  442\\
        7. Hybrid with rescore (lex. 0.5 sem. 0.5) & 25.20\% & 41.00\% & 442\\
        8. Hybrid with rescore (lex. 0.6 sem. 0.4) & 30.70\% & 42.30\% & 442\\
        \textbf{9. Hybrid with rescore (lex. 0.7 sem. 0.3)} &\textbf{ 30.80\%} & \textbf{42.50\%} & \textbf{442}\\
        \textbf{10. Hybrid with rescore (lex. 0.8 sem. 0.2)} & \textbf{ 30.80\%}& \textbf{42.50\%} &  \textbf{442} \\
        \textbf{11. Hybrid with rescore (lex. 0.9 sem. 0.1)}  &	 \textbf{30.80\%} & \textbf{42.50\%} &  \textbf{442}\\
        12. Lexical with stopwords removal & 28.70\% & 41.10\%  & 189 \\
        13. Lexical without stopwords removal & 28.50\% & 40.60\%  & 189 \\

\hline
        14. PubMed without MeSH Terms	& 9.20\% & 	15.30\% & 698 \\
        15. PubMed with MeSH Terms &	12.00\% &	19.10\% & 742	\\
\hline

    \end{tabular}%
}
\end{table}

The retrieval component was assessed under three configurations: (1) lexical-only, (2) semantic-only, and (3) a combination of lexical and semantic, with various weighting combinations tested for the hybrid approach.
For lexical retrieval, we experimented with and without stopword removal from the query. For semantic search, we tested three configurations:(1) semantic search with full embeddings, (2) compressed embeddings (using 8-bit quantization), and (3) compressed embeddings with re-scoring, where full embeddings were used for final ranking. Semantic search with full embeddings yielded an average response time of 30 seconds, making it impractical for real-world applications due to inefficiency; therefore, we excluded these results from our table.

The results summarized in Table \ref{tab_IR_performance} indicate that semantic retrieval alone yields suboptimal performance; however, integrating semantic and lexical retrieval substantially improves outcomes. Notable gains are observed when employing a hybrid search with a lexical weight of 0.1, which results in a 10.\% absolute increase in precision. Further improvement is seen at a lexical weight of 0.6, achieving a 5.5\% absolute increase. Increasing the lexical weight beyond 0.6 produces only marginal changes in performance. Notably, setting the lexical weight to 1, thereby excluding the semantic component, leads to a decline in precision.

Although three combinations of lexical and semantic weights yield the same top performance, we chose the one that gives the greatest emphasis on semantic search. This choice enables the system to capture better the deeper contextual relationships within biomedical texts, made possible by the contextual embeddings used in semantic search. Therefore, the configuration shown in the row 9 in Table \ref{tab_IR_performance} was selected for hybrid search in our system.

Additionally, we evaluated the performance of the PubMed search engine on the BioASQ dataset. When searching without MeSH terms, we obtained a P@10 of 9.2\% and a MAP@10 of 15.3\%. In contrast, when using MeSH terms, the performance improved to a P@10 of 12\% and a MAP@10 of 19.1\% (rows 14 and 15 in Table \ref{tab_IR_performance}). These results demonstrate that our retrieval approach outperforms the PubMed search engine, achieving significantly higher P@10 and MAP@10 scores, indicating a more effective ranking and retrieval of relevant abstracts for the BioASQ dataset. 

\subsection{Generative Component Evaluation}
\label{subsec:generative_component_standalone}

\subsubsection{Evaluation of references in generated answers}

To assess the impact of referenced QA fine-tuning, we evaluated the zero-shot results of Mistral-7B-Instruct-v0.2 (0-M2) and our fine-tuned version (M2) on the PQAref test set. The prompt used for 0-M2 model was:

\noindent
\qquad \\\fbox{
    \parbox{0.95\linewidth}{
Respond to the Instruction using only the information provided in the relevant abstracts under Abstracts. Reference the statements with the provided abstract\_id in brackets next to the statement (for example PUBMED:1235):\\\{instruction\}
       }
    }
\\

As part of its prompt template, M2 received the following prompt:

\noindent
\qquad \\\fbox{
    \parbox{0.95\linewidth}{Respond to the Instruction using only the information provided in the relevant abstracts in “‘Abstracts“‘ below. \\\{instruction\}
 }
    }
\\

In both cases, the \{instruction\} contained a concatenation of the user query and 10 retrieved abstracts with their PMIDs. We used default inference parameters, except for setting repetition\_penalty to 1.1 for M2 and adjusting max\_new\_tokens (or max\_tokens in zero-shot mode) to 1225 for both models.

We evaluated the characteristics of references on the PQAref test set (908 samples), focusing on the most common number of references per answer and the number of answers with no references. The results are presented in Table \ref{M2evaluations}.

Fine-tuning significantly changed the referencing behavior. 0-M2 generated 18.2\% of answers without references while still using the abstracts' information. After fine-tuning, M2 reduced this to only 5 answers (0.6\%), each stating that no abstracts were relevant or sufficient to answer the question. This demonstrates an improvement in task execution.

M2 most frequently generated answers with 3 references, showing a tendency to expand answers with additional, remotely related information. This differs from GPT-4 Turbo, which in the teacher role most often referenced only one abstract.

\subsubsection{Evaluation of Hallucinated PMIDs}

We evaluated whether the PMIDs in the generated answers fully matched those in the original context, ensuring there were no hallucinated PMIDs. Hallucination-free PMIDs are key since they provide the connection to the original abstracts, enabling us and the users to verify answers and access the source material. As a reference point, GPT-4 Turbo's answers in the PQAref dataset contained no hallucinated PMIDs. In contrast, 0-M2 hallucinated in case of 26 PMIDs, while fine-tuning reduced this number to only 3. The results of this evaluation are presented in the third row of Table \ref{M2evaluations}.

\begin{table}[h]
\centering
\caption{Evaluation of Mistral-7B-Instruct v0.2 in zero-shot mode (0-M2) and fine-tuned Mistral-7B-Instruct v0.2 (M2).}
\label{M2evaluations}
\resizebox{\linewidth}{!}{%
\begin{tabular}{c|l|>{\columncolor{gray!20}}c|c|c}
Samples                                                         & Evaluations                       & GPT-4 T    & 0-M2 & M2  \\ 
\hline
\multirow{4}{*}{\centering{908}} & Most references per answer   
& 1 (26.5\%) & 0  (18.2\%)            & 3 (19.6\%)             \\
\cline{2-5}

& Answers with no references & 2 (0.2\%) & 165 (18.2\%) & 5 (0.6\%) \\
\cline{2-5}
                                                                 & The  number of hallucinated PMIDs & 0 (0\%)                 & 26 (0.60\%)                       &\textbf{ 3 (0.08\%) }            \\
\cline{2-5}
                                                                 & BERTScore (F1)                    & 0.90                           & 0.84                                & \textbf{0.90}          \\ 
\hline
823                       & Missed most relevant abstract     & 1 (0.1\%)           & 185 (22.5\%)                     & \textbf{10 (1.2\%)}            \\ 
\hline
10                         & Recall for relevant abstracts     & 0.62        & 0.62                                  & \textbf{0.67}                \\ 
\hline                 
\end{tabular}
}
\raggedright


\end{table}

\subsubsection{Evaluation of Semantic Similarity}

We evaluated the similarity of the answer to the input. To assess how closely the answers aligned with the referenced abstracts, we calculated BERTScore \cite{zhangbertscore}, comparing each answer to the concatenation of the question and the abstracts cited within it. In cases where an answer included no references, we penalized the models by assigning a BERTScore of 0 to those answers. BERTScore for GPT-4 Turbo and M2 was at 0.90, indicating a high similarity between their answers and the referenced abstracts.
0-M2 scored slightly lower at 0.84, suggesting that, while it does generate similar content, it may lack the precision achieved through fine-tuning. The results of this evaluation are presented in the fourth row of Table \ref{M2evaluations}.

\subsubsection{Evaluation of Referencing Most Relevant Abstract}

To assess the relevance of the referenced abstracts, we used 823 samples of the PQAref dataset, which contain an abstract whose title matches the input question. These are the abstracts we consider the most relevant for that question. For those 823 samples, we evaluated how many times each of the models failed to reference at least this key abstract. When looking at the answers to 823 questions accompanied by their most relevant abstract, GPT-4 Turbo did not reference such an abstract in only one case, suggesting it served as a good referencing role model. 0-M2 missed the most relevant abstract in 22.5\% of answers, while M2 missed it in only 1.2\% of answers. This shows a stronger ability of the fine-tuned model to identify and extract the most relevant abstracts compared to its zero-shot version. The results of this evaluation are presented in the fifth row of Table \ref{M2evaluations}.

\subsubsection{Manual Evaluation of Recall for Relevant Abstracts}

We further performed a manual evaluation on a subset of 10 random samples, where we checked if the models referenced the abstracts deemed relevant by human annotators. To overcome variations in the number of relevant abstracts per document and document-specific characteristics, we considered all 100 abstracts (10 per question) collectively. We prioritized and calculated recall for the relevant abstracts for both models.

In terms of recall measured on the 10 manually reviewed samples, M2 exhibited a 0.05 increase compared to 0-M2 (from 0.62 to 0.67). For reference, the recall measured on the GPT-4 Turbo answers from the test set totaled 0.62. \hlchange{Qualitative analysis of these samples revealed distinct referencing behaviors. In the 10 observed examples, the fine-tuned model (M2) successfully referenced the "most relevant" abstract (the one providing a direct answer) in every single instance, whereas the zero-shot baseline failed to do so in 2 out of 10 cases.}

\hlchange{Furthermore, we observed that none of the models cited "completely irrelevant" abstracts. Instead, the general tendency was to cite "partially irrelevant" abstracts—documents that do not answer the specific question but offer valid, broad context. This suggests the models aim to provide comprehensive additional information rather than hallucinating links. We also noted sophisticated filtering behavior; for instance, in one case involving "donation," the models correctly excluded abstracts referring to blood donation when the context implied organ or tissue donation, demonstrating an ability to discern specific semantic scopes.}

Overall, the results suggest that fine-tuning improved content accuracy, PMID traceability, and the model’s ability to reference relevant source documents, making M2 a more reliable choice for our application.


\subsection{Verification Component Evaluation}
\label{subsec:aloneVEresults}

We conducted a comprehensive evaluation of the Verification Component across three settings: (1) in-domain testing using the transformed SciFact dataset; (2) out-of-domain testing on the HealthVer benchmark; and (3) comparison with GPT-4-based models in a zero-shot regime. These evaluations were designed to assess both the component’s claim verification performance and its robustness across domains and systems.

\subsubsection{In-Domain Evaluation}
\label{subsubsec:VC_in_domain}

The primary evaluation of the Verification Component was conducted on the test subset of the transformed SciFact dataset (see Section~\ref{subsec:VER} for dataset preparation). Among the four trained models, the best performance was achieved by a DeBERTa model fine-tuned with early stopping patience of 4. It reached a macro-averaged F1-score of 0.88 (Table~\ref{Taba1}).

The most notable challenge was accurate classification of the \textit{contradict} class. This class accounted for only 22\% of the training examples, contributing to reduced performance due to class imbalance. In contrast, the \textit{support} and \textit{no evidence} classes showed relatively higher classification accuracy. These results underscore the need for strategies such as data augmentation or class-balanced sampling to improve performance in underrepresented categories.

\begin{table*}[hbt!]
\small
\centering
\caption{The results of eight fine-tuned models 80\% of SciFact data, validated on 10\% of SciFact data, and tested on remaining 10\% of data}\label{Taba1}
\setlength{\tabcolsep}{2pt}

\begin{tabular}{@{}ccccccccccccccccccc@{}}
\toprule
\multicolumn{2}{c}{\multirow{2}{*}{}} & \multicolumn{4}{c}{RoBERTa L$_{SF}$} & \multicolumn{4}{c}{XLM RoBERTa L$_{SF}$} & \multicolumn{4}{c}{DeBERTa$_{SF}$} & \multicolumn{4}{c}{DeBERTa SQuAD$_{SF}$} \\ 
\cmidrule(l){3-6} \cmidrule(l){7-10} \cmidrule(l){11-14} \cmidrule(l){15-18}
\multicolumn{2}{c}{} & NE* & S & C & wa & NE & S & C & wa & NE & S & C & wa & NE & S & C & wa \\ 
\midrule
\multirow{4}{*}{3} & P & 0.71 & 0.55 & 0.00 & 0.48 & 0.83 & 0.69 & 0.54 & 0.71 & 0.83 & 0.86 & 0.85 & 0.84 & 0.86 & 0.90 & 0.82 & 0.87 \\
& R & 0.73 & 0.82 & 0.00 & 0.61 & 0.89 & 0.67 & 0.52 & 0.71 & 0.86 & 0.84 & 0.81 & 0.84 & 0.86 & 0.88 & 0.85 & 0.87 \\
& F1 & 0.72 & 0.66 & 0.00 & 0.53 & 0.86 & 0.68 & 0.53 & 0.71 & 0.84 & 0.85 & 0.83 & 0.84 & 0.86 & 0.89 & 0.84 & \textbf{0.87} \\
& Acc & \multicolumn{4}{c}{0.61} & \multicolumn{4}{c}{0.71} & \multicolumn{4}{c}{0.84} & \multicolumn{4}{c}{0.87} \\ 
\midrule
\multirow{4}{*}{4} & P & 0.85 & 0.75 & 0.67 & 0.77 & 0.75 & 0.76 & 0.71 & 0.74 & 0.88 & 0.90 & 0.88 & 0.89 & 0.82 & 0.91 & 0.88 & 0.87 \\
& R & 0.89 & 0.76 & 0.59 & 0.77 & 0.91 & 0.67 & 0.63 & 0.75 & 0.95 & 0.88 & 0.78 & 0.89 & 0.93 & 0.84 & 0.81 & 0.87 \\
& F1 & 0.87 & 0.76 & 0.63 & 0.77 & 0.82 & 0.71 & 0.67 & 0.74 & 0.91 & 0.89 & 0.82 & \textbf{0.88} & 0.87 & 0.88 & 0.85 & 0.87 \\
& Acc & \multicolumn{4}{c}{0.77} & \multicolumn{4}{c}{0.75} & \multicolumn{4}{c}{0.89} & \multicolumn{4}{c}{0.87} \\
\bottomrule
\multicolumn{19}{l}{* NE: no\_evidence, S: support, C: contradict, wa: weighted average, P: precision, R: recall,} \\
\multicolumn{19}{l}{  F1: F1 score, Acc: accuracy} \\

\end{tabular}
\end{table*}

\subsubsection{Out-of-Domain Evaluation}
\label{subsubsec:VC_out_domain}

To assess its generalization capacity, we evaluated the top DeBERTa model on the HealthVer dataset, a benchmark focused on fact-checking health-related claims using scientific evidence. Despite domain shifts and data format differences, the model achieved an F1-score of 0.44 and accuracy of 0.50, surpassing previous state-of-the-art results on this benchmark reported by \cite{sarrouti_evidence-based_2021}, where a BERT-base model fine-tuned on SciFact and evaluated on the HealthVer test set reached 0.36 F1 and 0.39 accuracy. Our model significantly outperforms the previous state-of-the-art on HealthVer, achieving absolute gains of 8 percentage points in F1-score and 11 percentage points in accuracy (Table~\ref{healthver}, DeBERTa$_{SF-80}$).

We further explored the impact of additional training data by expanding the fine-tuning set to 90\% of the transformed SciFact dataset. The retrained model (DeBERTa$_{SF-90}$) achieved an additional improvement of 4 percentage points in both F1-score and accuracy, confirming that increased training volume can enhance model robustness in downstream biomedical claim verification tasks. This final version was selected as the core model for our Verification Component and is publicly available via HuggingFace\footnote{\url{https://huggingface.co/MilosKosRad/TextualEntailment_DeBERTa_preprocessedSciFACT}}.

\begin{table}[hbt!]

\caption{Results of the DeBERTa model fine-tuned on the 80\% and 90\% of the SciFact dataset end evaluated on the HealthVer test set.}
\label{healthver}
\begin{tabular}{@{}cccccccccc@{}}
\toprule
          & \multicolumn{5}{c}{DeBERTa$_{SF-80}$}                    & \multicolumn{4}{c}{DeBERTa$_{SF-90}$}                    \\
\cmidrule{2-5} \cmidrule{7-10}
          & NE & S & C & wa & & NE & S & C & wa \\
\cmidrule{2-5} \cmidrule{7-10}
P & 0.46        & 0.70    & 0.66       & 0.60    &     & 0.47        & 0.67    & 0.69       & 0.59         \\
R    & 0.94        & 0.25    & 0.15       & 0.50  &       & 0.88        & 0.29    & 0.27       & 0.52         \\
\cmidrule{2-5} \cmidrule{7-10}
F1  & 0.62        & 0.37    & 0.24       & \textbf{0.44}    &     & 0.61        & 0.40    & 0.39       & \textbf{0.48}         \\
\cmidrule{2-5} \cmidrule{7-10}
Acc  & \multicolumn{4}{c}{0.50}                          & \multicolumn{4}{c}{0.52}\\
\bottomrule
\end{tabular}

\end{table}

\subsubsection{Comparison with GPT-4 Models}
\label{subsubsec:CV_vs_gpt4}

To contextualize the performance of our fine-tuned verification model against powerful general-purpose models, we evaluated GPT-4, GPT-4 Turbo, and GPT-4o in a zero-shot setting. Each model was prompted with the same 122 examples used for the in-domain test set (i.e., the same 10\% of our transformed SciFact dataset). We standardized parameters across models (e.g., temperature set to 0, max tokens set to 350) to minimize variability and maximize determinism.
The prompt we used was as follows:

\noindent
\qquad \\\fbox{
    \parbox{0.95\linewidth}{
Critically asses whether the statement is supported, contradicted or there is no evidence for the statement in the given abstract. Output SUPPORT if the statement is supported by the abstract. Output CONTRADICT if statement is in contradiction with the abstract and output NO\_EVIDENCE if there is no evidence for the statement in the abstract.
       }
}
\\

Our DeBERTa model consistently outperformed all GPT-4 models in both F1-score and accuracy (Table~\ref{gpt4}), confirming that domain-specific fine-tuning yields superior results in tasks involving nuanced biomedical claim verification. Moreover, our open-source solution ensures transparency, auditability, and adaptability—factors that are critical in healthcare, scientific, and regulatory domains.

\begin{table*}[hbt!]
\small
\centering
\caption{Comparison of the DeBERTa$_{SF}$ model with GPT-4 models.}
\label{gpt4}
\setlength{\tabcolsep}{3.5pt}

\begin{tabular}{llllllllllllllllllll}
\toprule
    & \multicolumn{4}{c}{DeBERTa$_{SF}$}    & \multicolumn{1}{c}{} & \multicolumn{4}{c}{GPT-4}                   & \multicolumn{1}{c}{} & \multicolumn{4}{c}{GPT-4 Turbo} & \multicolumn{1}{c}{} & \multicolumn{4}{c}{GPT-4o} \\ \cmidrule{2-5} \cmidrule{7-10} \cmidrule{12-15} \cmidrule{17-20}
    & NE   & S    & C    & wa            &                      & NE   & S    & C             & wa            &                      & NE     & S      & C     & wa    &                      & NE    & S    & C    & wa   \\ \cmidrule{2-5} \cmidrule{7-10} \cmidrule{12-15} \cmidrule{17-20}
P   & 0.88 & 0.90 & 0.88 & 0.89          &                      & 0.85 & 0.77 & 0.84          & 0.82          &                      & 0.93   & 0.81   & 0.65  & 0.82  &                      & 0.72  & 0.91 & 0.74 & 0.80 \\
R   & 0.95 & 0.88 & 0.78 & 0.89          &                      & 0.80 & 0.94 & 0.59          & 0.81          &                      & 0.64   & 0.92   & 0.81  & 0.80  &                      & 0.89  & 0.80 & 0.63 & 0.80 \\ \cmidrule{2-5} \cmidrule{7-10} \cmidrule{12-15} \cmidrule{17-20}
F1  & 0.91 & 0.89 & 0.82 & \textbf{0.88} & \textbf{}            & 0.82 & 0.85 & 0.70 & \textbf{0.81} & \textbf{}            & 0.76   & 0.86   & 0.72  & 0.79  &                      & 0.80  & 0.85 & 0.68 & 0.79 \\ \cmidrule{2-5} \cmidrule{7-10} \cmidrule{12-15} \cmidrule{17-20}
Acc & \multicolumn{4}{c}{0.89}           & \multicolumn{1}{c}{} & \multicolumn{4}{c}{0.81}                    & \multicolumn{1}{c}{} & \multicolumn{4}{c}{0.80}        & \multicolumn{1}{c}{} & \multicolumn{4}{c}{0.80}  \\ \bottomrule
\end{tabular}
\end{table*}

\subsubsection{Error Analysis}
\label{subsubsec:VC_errors}

\begin{figure}[!h]
\centering
\includegraphics[width=0.75\linewidth]{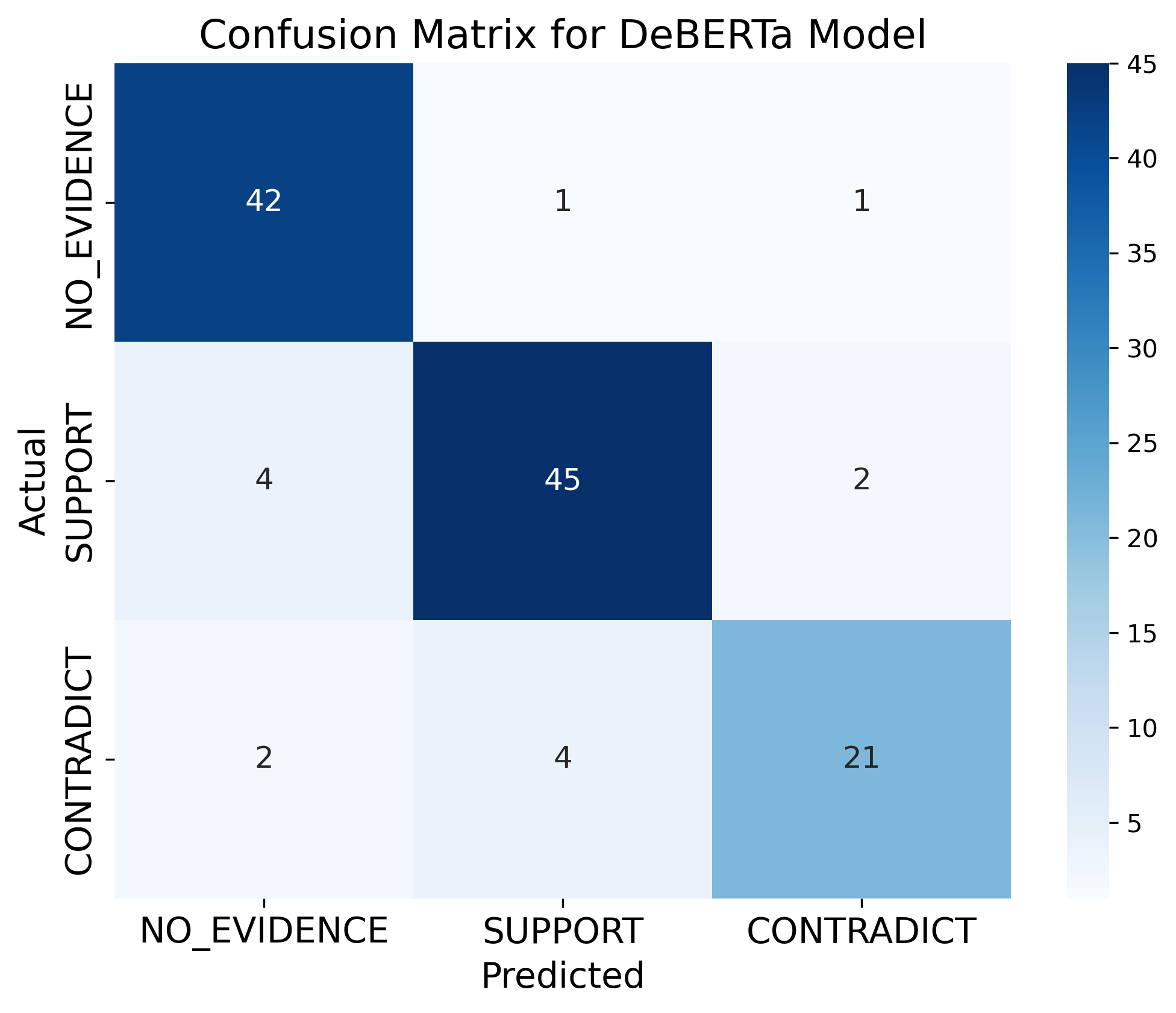}
\caption{Confusion matrix for DeBERTa$_{SF}$ model.}
\label{fig:confusionDeBERTa}
\end{figure}

A comprehensive error analysis was conducted on the in-domain test set to scrutinize the limitations of our top-performing model, DeBERTa$_{SF}$. We analyzed all 14 misclassified instances out of the 122 claim-abstract pairs. As illustrated in Figure~\ref{fig:confusionDeBERTa}, while the model exhibits commendable performance in the \textit{No Evidence} and \textit{Support} classes, the \textit{Contradict} class remains a focal point for improvement.

\hlchange{In the \textit{Support} class, misclassifications into \textit{No Evidence} primarily stemmed from the model's inability to reconcile dense numerical data or recognize abbreviations. Conversely, when \textit{Support} claims were mislabeled as \textit{Contradict}, the error was often attributable to the semantic complexity of clinical trial data. Specifically, complex immunology terminology and the misalignment of specific time frames between claim and evidence.}

\hlchange{However, the highest misclassification rate was observed in the \textit{Contradict} class. Out of 27 contradictory examples, 6 were misclassified (4 as \textit{Support}, 2 as \textit{No Evidence}). These errors highlight two distinct failure modes:}

\begin{enumerate}
\item \textbf{Surface-level Similarity (\textit{Contradict} $\rightarrow$ \textit{Support}):} \hlchange{The model often relied on lexical overlap (e.g., matching keywords) rather than semantic logic. This suggests the model may overlook clear contradictions when misled by generalizations or shared phrasing.}
\item \textbf{Implicit Negation (\textit{Contradict} $\rightarrow$ \textit{No Evidence}):} \hlchange{The absence of explicit negation in training data made it difficult for the model to detect subtle contradictions, particularly when embedded in experimental variations or implicit semantic inversions.}
\end{enumerate}

\hlchange{To better understand these failures, we examined two representative cases where the model struggled with nuanced scientific reasoning.}

\hlchange{First, the claim \textit{"The genomic aberrations found in matasteses are very similar to those found in the primary tumor"} was misclassified as \textit{Support}, despite evidence indicating divergent evolutionary trajectories. Several factors contributed to this:}
\begin{itemize}

\item \hlchange{\textbf{Lexical Noise:} The term "metastases" was misspelled in the claim ("matasteses") but correct in the text, introducing noise.}

\item \hlchange{\textbf{Evolutionary Complexity:} The claim implies a linear, straightforward similarity. In contrast, the evidence describes a parallel evolutionary process involving distinct genetic signatures and dissemination routes—a distinction the model failed to capture.}
\item \hlchange{\textbf{Semantic Overlap:} The high overlap of genetic terminology likely triggered a "support" prediction based on keyword matching alone.}

\end{itemize}

\hlchange{Second, the claim \textit{"The most prevalent adverse events to Semaglutide are cardiovascular"} was incorrectly labeled as \textit{No Evidence}. The evidence actually contradicted this by citing gastrointestinal events as the most common side effect. The model's failure here was driven by:}
\begin{itemize}
\item \hlchange{\textbf{Information Overload:} The evidence contained extensive details on participant demographics, dosage specifics (e.g., 2.5 mg vs 5 mg), and diverse efficacy outcomes, overwhelming the inference mechanism.}
\item \hlchange{\textbf{Implicit Contradiction:} The contradiction was not explicitly stated as a negation of cardiovascular events; rather, it was implicit in the emphasis on gastrointestinal events.}
\item \hlchange{ \textbf{Numerical Density:} The presence of various numerical data points (HbA1c levels, body weight changes) appeared to obscure the relevant semantic signal regarding adverse event prevalence.}
\end{itemize}

These observations suggest three critical directions for future work: (1) targeted data augmentation to include more contradictory examples with high lexical overlap; (2) increased exposure to domain-specific numerical and clinical data to improve robustness against "numerical noise"; and (3) adversarial evaluation focusing on claims with subtle semantic inversions and implicit negations.

\subsection{End-to-end System Evaluation}

After identifying the best-performing configurations for each component of our pipeline --- Information Retrieval Component (IR), Generative Component (GC), and Verification Component (VC) --- we conducted an end-to-end evaluation of the complete VerifAI system. The aim was to assess how effectively the integrated pipeline performs the full task of biomedical question answering, including referenced evidence retrieval and claim-level factual verification.

To ensure consistency and comparability across components, we selected a subset of 182 questions from the BioASQ dataset, each associated with exactly 10 relevant abstracts - the same number used during GC fine-tuning. After excluding three samples with missing abstracts and one with an incomplete gold answer, the final evaluation set comprised 178 questions.

\subsubsection{Information Retrieval Component}

To evaluate the IR component in an end-to-end setting, we first assessed its ability to retrieve abstracts that matched the gold standard abstracts provided by BioASQ. We used the same evaluation metrics as in Section~\ref{subsec:IR_component_stamdalone}, focusing on Precision@10 (P@10) and Mean Average Precision at 10 (MAP@10). As shown in Table~\ref{table:IR_evaluation_10Abstract}, our system achieves a P@10 of 23.7\% and a MAP@10 of 42.7\%, indicating that relevant documents are ranked reasonably well within the top 10 results. These scores are slightly lower in P@10 but higher in MAP than in the standalone IR evaluation, suggesting consistent but dataset-dependent behavior.

\begin{table}[h]
    \centering
    \caption{Our IR and PubMed search engine performance evaluation on the BioASQ subset (exactly 10 PubMed abstracts per question).
    }
    \label{table:IR_evaluation_10Abstract}
    \resizebox{\columnwidth}{!}{%
    \begin{tabular}{lcccc}
\hline
         & P@10 & MAP@10 & time [ms] \\
\hline
        1. Semantic without rescore & 12.90\% & 32.20\%  & 190  \\
        2. Semantic with rescore & 13.00\% & 32.70\% & 300\\
        3. Hybrid with rescore (lex. 0.1 sem. 0.9) & 16.00\% & 35.80\% & 451\\
        4. Hybrid with rescore (lex. 0.2 sem. 0.8) & 16.00\% & 35.70\% & 476\\
        5. Hybrid with rescore (lex. 0.3 sem. 0.7) & 16.00\% & 35.50\% & 443\\
        6. Hybrid with rescore (lex. 0.4 sem. 0.6) & 16.00\% & 35.90\% & 448\\
        7. Hybrid with rescore (lex. 0.5 sem. 0.5) & 16.40\% & 41.00\% & 401\\
        8. Hybrid with rescore (lex. 0.6 sem. 0.4) & 23.60\% & 42.30\% & 425\\
        \textbf{9. Hybrid with rescore (lex. 0.7 sem. 0.3)}  &\textbf{23.70\%} & \textbf{42.70\%} & \textbf{401}\\
        \textbf{10. Hybrid with rescore (lex. 0.8 sem. 0.2)}  & \textbf{ 23.70\%} & \textbf{42.70\%} & \textbf{448} \\
        \textbf{11. Hybrid with rescore (lex. 0.9 sem. 0.1)}  &	\textbf{ 23.70\%} &\textbf{42.70\%} & \textbf{451}\\
        12. Lexical with stopwords removal & 22.60\% & 41.80\%  & 173 \\
        13. Lexical without stopwords removal & 21.80\% & 40.40\%  & 230 \\

\hline
        14. PubMed without MeSH Terms	&5.50 \% &12,20\% & 1060 \\
        15. PubMed with MeSH Terms &	7.70\% & 16.60\% & 	1124\\
\hline

    \end{tabular}%
}
\end{table}

To further contextualize retrieval effectiveness, Table~\ref{tab:matching_all_by_query_type} breaks down how many abstracts matched per question. More than 76\% of the questions (136 out of 178) had at least one matched abstract. Only one question had all 10 abstracts retrieved correctly, while 42 had no overlaps. These retrieval statistics suggest that while perfect matching is rare, most questions are accompanied by at least partially relevant content retrieved by our hybrid IR strategy.

\begin{table}[t]
    \centering
    \caption{Number of questions per type grouped by the number of matching abstracts between our IR system and BioASQ} 
    \begin{tabular}{c ccccc}
        \toprule
        \textbf{Matching Abstracts} & \textbf{Factoid} & \textbf{List} & \textbf{Summary} & \textbf{Yes/No} & \textbf{Total}\\
        \midrule
        0  & 11 & 13 & 9 & 9 & 42\\
        1  & 13 & 9 & 7 & 11 & 40\\
        2  & 6& 3 & 7 & 7 & 23\\
        3  & 2 & 6 & 3 & 5 & 16\\
        4  & 6 & 2 & 7 & 7 & 22\\
        5  & 4 & 2 & 6 & 4 & 16\\
        6  & 5 & 1 & 2 & 4 & 12\\
        7  & 0 & 0 & 0 & 1 & 1\\
        8  & 2 & 0 & 0 & 3  & 5 \\
        9  & 0 & 0 & 0 & 0 & 0 \\
        10 & 0 & 0 & 1 & 0 & 1\\
        \midrule
        Total & 49 & 36 & 42 & 51 & 178\\
        \bottomrule
    \end{tabular}
    
    \label{tab:matching_all_by_query_type}
\end{table}

\subsubsection{Generative Component}

We evaluated the GC outputs under two different IR inputs: (1) using the 10 gold-standard abstracts provided by BioASQ, and (2) using the top 10 abstracts retrieved by our IR component. In both settings, our GC model (M2) generated answers conditioned on these abstracts, and the outputs were compared against the human-written ideal answers in BioASQ.

For the questions in which BioASQ offers more than one ideal answer, we used GPT-4 Turbo to combine the answers into one. The generated and reference answers were then evaluated by GPT-4 Turbo using a structured prompt, which asked three things: (1) whether the generated answer reaches the same conclusion as the reference (YES/NO), (2) whether it includes all the information from the reference (YES/NO), and (3) if not, what percentage of the information is included. The full prompt is provided below:

\noindent
\qquad \\\fbox{
    \parbox{0.95\linewidth}{Compare the sample answer to the ideal answer. The sample answer can be more detailed as long as it contains all the information from the ideal answer. Include this information in your comparison: 1. Do the answers come to the same general conclusion? Answer with YES or NO, under the variable SAME\_CONCLUSION. 2. Does the sample answer contain all the information covered by the ideal answer? Answer with YES or NO under the variable ALL\_INFO. If the answer is NO to any of these questions, say what exactly is missing in the sample answer. Ignore the PMIDs in the sample answer. Calculate the percentage of crucial information from the ideal answer that is covered in the sample answer (with max of 100\%) and state it under the variable PERC\_IDEAL. Explain your calculation.}
    }
\\

BioASQ categorizes questions into four types that reflect different reasoning and generation demands: yes/no (binary judgment), factoid (a specific fact or entity), list (multiple entities), and summary (a short paragraph synthesizing relevant information). These types serve as a useful lens for analyzing model performance across varying answer formats. Table~\ref{tab:endtoendgeneval} presents both per-type and overall results. When using the gold-standard abstracts, M2’s answers reached the same conclusion as the reference in approximately 81\% of cases and included about 75\% of the reference information. However, only 27 out of 178 answers referenced all 10 abstracts, underscoring the difficulty of comprehensive citation.

\begin{table}[h]
\centering
\caption{End-to-end evaluation of the generative component.}
\resizebox{\linewidth}{!}{%
\begin{tabular}{cccccc}
\hline
Input & Type & Samples & SAME\_CONCLUSION & ALL\_INFO & PERC\_IDEAL \\
\hline
\multirow{5}{*}{\rotatebox{90}{BioASQ}} 
& Yes/No & 51 & 82.35\% & 52.94\% & 74.43\% \\
& Factoid & 49 & \textbf{85.71\%} & \textbf{59.18\%} & 74.72\% \\
& Summary & 42 & 73.81\% & 40.48\% & 69.59\% \\
& List & 36 & 83.33\% & 44.44\% & \textbf{80.36\%} \\
\cline{2-6}
& all & 178 & 81.46\% & 50\% & 74.57\% \\
\hline
\multirow{5}{*}{\rotatebox{90}{\makecell{BioASQ\\our IR}}}
& Yes/No & 51 & \textbf{82.35\%} & \textbf{49.02}\% & \textbf{73.85\%}\\
& Factoid & 49 & 73.47\% & 46.94\% & 72.07\% \\
& Summary & 42 & 69.05\%& 28.57\% & 64.56\%\\
& List & 36 & 52.78\% & 22.22\% & 57.18\%\\
\cline{2-6}
& all & 178 & 70.79\% & 38.20\% & 67.62\% \\
\hline
\end{tabular}
}
\label{tab:endtoendgeneval}
\end{table}

When using abstracts retrieved by our IR component instead of the gold-standard set, scores were generally lower but followed similar trends across question types. The yes/no questions remained the most robust, with no drop in conclusion agreement, suggesting that the GC model can often infer binary answers correctly even with slight evidence variations. In contrast, list questions showed the steepest performance decline, which aligns with their stronger dependency on exact document matches for accurate item enumeration. This is further supported by the retrieval statistics, where list questions had the highest proportion of samples (13 out of 36) with zero matching abstracts (see Table~\ref{tab:matching_all_by_query_type}).

These results show that the quality of retrieved abstracts significantly affects the performance of the generative model – especially for question types that require multiple specific facts, such as lists or factoids. This underscores the importance of retrieval systems that can return as many relevant documents as possible, ensuring that the generator has sufficient information to produce accurate and complete answers.

\subsubsection{Verification Component}

To evaluate the end-to-end performance of the Verification Component, we again used the 178 BioASQ questions with two retrieval setups: (1) gold-standard abstracts from BioASQ and (2) abstracts retrieved by our IR component. Each generated answer was linked to specific PubMed abstracts via citations (PMIDs), allowing us to form claim–evidence pairs by associating each claim with its referenced abstract. These pairs were passed to our best-performing VE model (DeBERTa$_{SF-90}$, see Section~\ref{subsec:aloneVEresults}) for classification into one of three categories: \textit{support}, \textit{contradict}, or \textit{no evidence }.

Since BioASQ does not include such NLI labels, we generated reference labels for the question-abstract pairs using GPT-4 and GPT-4o, allowing us to compare our model’s output against two variants of a leading large language model in a real-world setting.


\begin{table*}[hbt!]
\small
\centering
\caption{Performance evaluation of models judged by GPT-4 and GPT-4o on retrieval from BioASQ and by pipeline's IR component.}
\label{tab:VEcomparison}
\setlength{\tabcolsep}{3.5pt}
\resizebox{\linewidth}{!}{%
\begin{tabular}{llllllllllllllllllll}
\toprule
    & \multicolumn{4}{c}{GPT-4 (BioASQ)}    & \multicolumn{1}{c}{} & \multicolumn{4}{c}{GPT-4o (BioASQ)} & \multicolumn{1}{c}{} & \multicolumn{4}{c}{GPT-4 (ourIR)} & \multicolumn{1}{c}{} & \multicolumn{4}{c}{GPT-4o (ourIR)} \\ 
    \cmidrule{2-5} \cmidrule{7-10} \cmidrule{12-15} \cmidrule{17-20}
    & NE   & S    & C    & wa            &                      & NE   & S    & C    & wa            &                      & NE   & S    & C    & wa    &                      & NE   & S    & C    & wa   \\ 
    \cmidrule{2-5} \cmidrule{7-10} \cmidrule{12-15} \cmidrule{17-20}
P   & 0.89 & 0.77 & 0.17 & 0.83          &                      & 0.83 & 0.81 & 0.19 & 0.81          &                      & 0.88 & 0.83 & 0.11 & 0.85  &                      & 0.82 & 0.90 & 0.08 & 0.86 \\
R   & 0.75 & 0.88 & 0.42 & 0.81          &                      & 0.82 & 0.80 & 0.33 & 0.81          &                      & 0.80 & 0.87 & 0.44 & 0.83  &                      & 0.90 & 0.79 & 0.22 & 0.84 \\
\cmidrule{2-5} \cmidrule{7-10} \cmidrule{12-15} \cmidrule{17-20}
F1  & 0.81 & 0.83 & 0.24 & 0.81          &                      & 0.83 & 0.81 & 0.24 & 0.81          &                      & 0.84 & 0.85 & 0.18 & 0.84  &                      & 0.86 & 0.84 & 0.12 & 0.85 \\
\cmidrule{2-5} \cmidrule{7-10} \cmidrule{12-15} \cmidrule{17-20}
Acc & \multicolumn{4}{c}{0.81}           & \multicolumn{1}{c}{} & \multicolumn{4}{c}{0.81}        & \multicolumn{1}{c}{} & \multicolumn{4}{c}{0.83}        & \multicolumn{1}{c}{} & \multicolumn{4}{c}{0.84}  \\ 
\bottomrule
\end{tabular}}
\end{table*}

As seen in Table~\ref{tab:VEcomparison}, the VC model achieved 81\% accuracy using BioASQ abstracts and up to 84\% accuracy using our IR abstracts, suggesting that our retrieval pipeline may provide more contextually appropriate evidence for entailment classification. In both scenarios, the model reliably identified \textit{support} and \textit{no evidence } cases (F1 scores of 0.81–0.86), while the \textit{contradict} category remained the most difficult, with lower F1 scores (0.12–0.24).

Interestingly, GPT-4 showed better recall for contradictions, while GPT-4o demonstrated higher precision and slightly better overall F1. Despite these strengths, both GPT models underperformed compared to our VC model in standalone evaluations (Section~\ref{subsubsec:CV_vs_gpt4}), and these trends were largely consistent in this end-to-end setting.

Together, these results validate the robustness and complementarity of our pipeline. While each component faces specific challenges --- such as retrieval coverage, abstract-based summarization, or contradiction detection --- the integration of specialized modules trained and tuned for their respective subtasks proves highly effective in delivering fact-checked, referenced answers to biomedical questions.

\section{Discussion}
\label{sec:discussion}
Our end-to-end evaluation demonstrates that VerifAI successfully integrates information retrieval, generative QA, and verification into a coherent pipeline for biomedical question answering. Each component plays a critical role: hybrid retrieval boosts document relevance, which directly impacts both the factual completeness of generated answers and the accuracy of claim verification. This interdependence highlights the importance of optimizing all components jointly, rather than in isolation.

The generative model (M2) showed strong performance, particularly when supported by high-quality retrieved evidence. While its outputs captured the core conclusions in over 80\% of cases and preserved a substantial portion of reference information, performance was predictably sensitive to retrieval quality. This effect was most evident for list-based questions, which require broader evidence coverage and exact entity matching.

\hlchange{A key design decision in VerifAI is the decoupling of generation and verification. While the generative component relies on parametric knowledge (prone to hallucinations and outdated facts), the verification component operates strictly as a NLI engine. By framing verification as a discriminative task between a provided premise (retrieved abstract) and a hypothesis (claim), we mitigate the risk of "knowledge hallucinations" common in LLMs. The verifier does not need to "know" the fact; it only needs to assess logical entailment within the provided context window. This allows fine-tuned SLMs like DeBERTa to outperform much larger models like GPT-4 on benchmarks such as HealthVer, as they are optimized for logical reasoning rather than open-ended generation.}

A key insight from the verification component is that retrieval quality affects entailment classification. Using abstracts retrieved by our hybrid IR system led to higher verification accuracy than using gold-standard BioASQ abstracts --- suggesting that VerifAI’s retrieval strategy may be more aligned with the inferential needs of NLI models. Moreover, our verification engine consistently outperformed GPT-4 models in both standalone and end-to-end settings, confirming its robustness and domain-specific effectiveness.

\begin{figure*}[h!]
    \centering
    \includegraphics[width=0.8\linewidth]{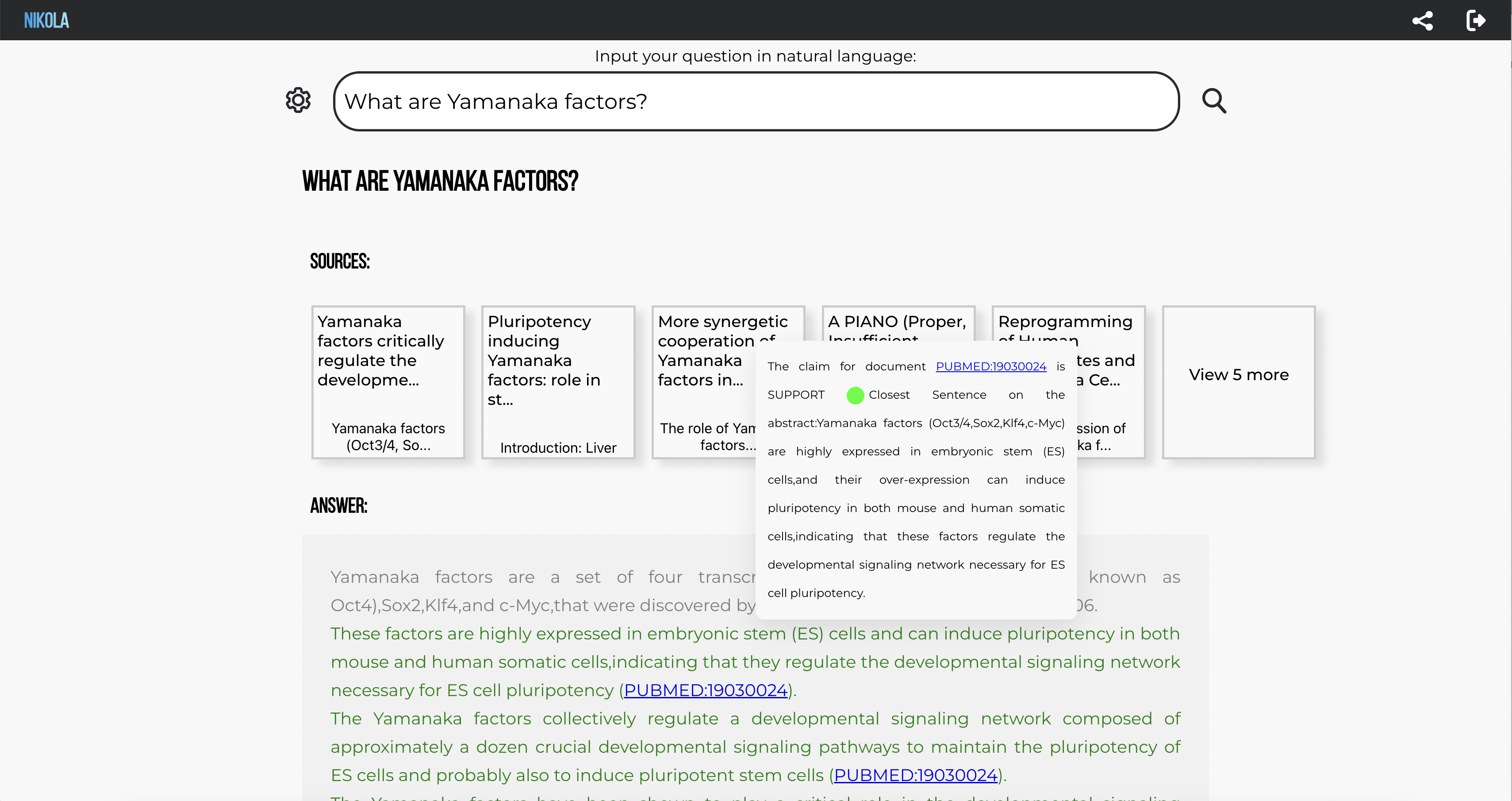}
    \caption{\hlchange{Screenshot of the VerifAI user interface. Users enter a biomedical question in the input box; the system then displays (1) \textit{Sources}, i.e., the combined output of lexical and semantic retrieval, and (2) \textit{Answer}, produced by the generative component, where each sentence is color-coded according to the verification engine’s label. Hovering over a sentence reveals its predicted attribution class, the linked PubMed reference (if available), and the semantically closest sentence from the cited abstract.}}
    \label{fig:placeholder}
\end{figure*}

These results underscore the value of designing AI pipelines where generation is grounded in traceable evidence and independently verifiable claims. While handling contradictions remains a challenge --- as it is for all NLI systems in complex domains --- the system’s performance offers a strong foundation for trustworthy, reference-aware biomedical QA.

\hlchange{While the modular architecture improves grounding, it introduces a latency trade-off. The sequential execution of generation followed by claim-level verification (where each claim triggers a separate NLI inference pass) inevitably increases total response time compared to standard RAG, even when the NLI inference passes are triggered in parallel for all generated claims. However, for high-stakes domains like biomedicine, we argue that this computational cost is a necessary investment for verifiable reliability.}

\hlchange{Beyond the technical accuracy of the verification engine, VerifAI's user interface is designed to facilitate rapid, transparent validation of generated claims. The system employs a color-coded text scheme where supported sentences are rendered in dark green font, partially supported claims appear in dark yellow/orange, contradictions are highlighted in red, and unreferenced sentences are displayed in dark gray (see Figure}~\ref{fig:placeholder}\hlchange{). This immediate visual feedback allows users to quickly identify the trustworthiness of each statement without needing to parse technical classification labels. More importantly, hovering over any sentence reveals three critical pieces of information: the entailment class assigned by the verification engine, the referenced PubMed abstract (if applicable), and the semantically closest sentence from that abstract. This hover feature is particularly valuable for domain experts who wish to validate the system's reasoning: by presenting the exact source sentence alongside the generated claim, users can independently assess whether the entailment classification is justified. This design philosophy aligns with principles of explainable AI, where transparency and user agency are prioritized over black-box decision-making} \cite{arrieta2020explainable}. \hlchange{In practice, this means that even when the verification engine makes a mistake — the user is equipped with the raw evidence needed to override or contextualize the system's judgment.}

Future work will focus on fine-grained improvements in contradiction detection, further modular evaluations in other high-stakes domains, and integration of human expert feedback to assess real-world applicability and reliability.

\section{Comparison with Other Systems}
\label{sec:comparison}

While several platforms exist for scientific or general-purpose question answering, VerifAI remains unique in combining retrieval, generation, and claim verification in a single, domain-adapted pipeline. Unlike web-based tools such as Perplexity or You.com, which retrieve open-domain content and generate answers without factual validation, VerifAI grounds its responses in trusted biomedical literature and verifies generated claims against retrieved evidence.

Tools like Scite and Consensus offer structured scientific search and citation-based relevance, and Scite additionally provides RAG-style generation with inline citations. However, these systems do not incorporate explicit verification of claims, nor do they offer flexible model customization or dataset indexing. VerifAI, in contrast, provides a fully modular, open-source framework that enables fine-tuning across retrieval, generation, and NLI verification components --– making it adaptable to different domains and use cases.

\begin{table*}[h]
\centering
\scriptsize
\setlength{\tabcolsep}{10pt}
\caption{Comparison of VerifAI with Other AI-Driven Systems (Shortened)}
\label{tab:verifai_comparison_short}
\begin{tabular}{
    l
    >{\raggedright\arraybackslash}p{1.3cm}
    >{\raggedright\arraybackslash}p{1.3cm}
    >{\raggedright\arraybackslash}p{1.9cm}
    >{\raggedright\arraybackslash}p{1.9cm}
    >{\raggedright\arraybackslash}p{2.3cm}
    c
    c
    }
\toprule
System & Focus & Retrieval & Citations & Verification & Model & Custom & Open \\
\midrule
VerifAI    & Biomed QA & Lex+Sem & Yes (PubMed)             & Yes                 & Mistral-7B                       & Yes & Yes \\
\addlinespace[0.8em]
Elicit     & Lit. Review & Paper IR & Yes (Papers)            & No                        & GPT4                            & No  & No  \\
\addlinespace[0.8em]
Perplexity & General & Web & Yes (URLs)                  & No                        & Own (Llama-based: Sonnar) / Comm. (e.g. GPT4) & No  & No  \\
\addlinespace[0.8em]
You.com    & General & Web & Yes (URLs)                  & No                        & Comm. (GPT4, Claude 3, etc.)     & No  & No  \\
\addlinespace[0.8em]
Scite      & Scientific & Hybrid IR & Yes (Citations)        & Yes                       & o3-mini                         & No  & No  \\
\addlinespace[0.8em]
Consensus  & Sci. Search & Academic & Yes (Papers)           & No                        & GPT4                            & No  & No  \\
\bottomrule
\end{tabular}
\end{table*}

Compared to these systems, VerifAI offers a distinct advantage by tightly integrating domain-specific retrieval, controlled generation, and explicit claim verification --- all within a fully modular, open-source pipeline. This design not only ensures higher factual consistency but also provides researchers and institutions with full control over the indexed corpus, model components, and evaluation workflows. Such flexibility is critical in biomedical contexts, where trust, reproducibility, and domain alignment are essential. As shown in Table~\ref{tab:verifai_comparison_short}, no other platform offers this level of end-to-end transparency and adaptability.

\section{Generalizing VerifAI for Domain-Agnostic Verifiable Search and Extensibility to Other domains}
\label{sec:generalization}

Although VerifAI was originally optimized for biomedical question answering grounded in PubMed abstracts, its architecture has evolved into a flexible, domain-agnostic framework. \hlchange{The system's modular design---separating Information Retrieval (IR), Generative Component (GC), and Verification Component (VC)---allows for seamless adaptation to other high-stakes fields such as law, finance, or policy making. Unlike end-to-end black-box models, each module interacts through standardized interfaces (queries, ranked documents, cited claims, and entailment labels), allowing practitioners to swap individual components without redesigning the entire pipeline.}

In this section, we outline the strategic considerations for adapting VerifAI’s three core layers to new domains.

\subsection{Generative Strategy: Prompting vs. Fine-tuning}

\hlchange{A critical decision in cross-domain adaptation is the choice of the generative model.}
\begin{itemize}
    \item \hlchange{\textbf{General-Purpose Prompting:} For domains with limited training data, utilizing a large, general-purpose LLM (e.g., GPT-4, Llama 3) via API is a viable starting point. VerifAI natively supports OpenAI-compatible endpoints (e.g., vLLM, Ollama, Nvidia NIM, OpenAI API), lowering the barrier to entry. However, this approach necessitates rigorous prompt engineering to enforce strict citation formatting and may be less consistent in adhering to domain-specific stylistic norms.}
    \item \hlchange{\textbf{Task-Specific Fine-tuning:} As demonstrated with our Mistral-7B model, fine-tuning offers superior control over output structure and citation behavior. This strategy is particularly beneficial when the target domain requires a specific answer format (e.g., a "holding" in legal memos), when context windows must be optimized, or when data privacy concerns necessitate smaller, locally deployed models.}
\end{itemize}

\subsection{IR Layer Adaptation}
\hlchange{The hybrid retrieval logic (lexical + semantic) is universally applicable, but the underlying artifacts must be domain-aligned.}
\begin{itemize}
    \item \hlchange{\textbf{Corpus \& Metadata:} The retrieval target must be swapped (e.g., replacing PubMed with the SEC EDGAR database or internal enterprise wikis). VerifAI currently supports indexing diverse formats including PDF, DOCX, PPTX, TXT, Markdown (MD), and EPUB.}
    \item \hlchange{\textbf{Embedding Models:} While general-purpose embeddings work well, domain-specific embedding models (e.g., LawBERT, FinBERT) should be employed to capture specialized semantic relationships that general models might miss.}
\end{itemize}

\subsection{Verification Layer Adaptation}
\hlchange{The verification component is the "trust layer" of the system. While the NLI formulation (Support/Contradict/No Evidence) is standard, the definition of contradiction varies by field.}
\begin{itemize}
    \item \hlchange{\textbf{Semantic Nuance:} In biomedicine, contradictions often involve experimental conditions or numerical data. In law, contradictions may hinge on jurisdictional overruling or procedural distinctions.}
    \item \hlchange{\textbf{Training Strategy:} Transferring the VC requires either (1) prompting a reasoning-strong model with domain-specific guidelines or (2) fine-tuning a BERT-style model on a domain-specific NLI dataset. We suggest starting with a strong general NLI model and progressively fine-tuning on domain-specific "hard negatives" to improve sensitivity to subtle contradictions.}
\end{itemize}

\hlchange{By following this modular adaptation strategy, VerifAI can serve as a foundational template for creating verifiable, citation-backed search engines in any domain where accuracy and source transparency are paramount. At Bayer, for instance, a customized version of VerifAI has been successfully deployed to search internal regulatory document databases.}

\section{Conclusion}
\label{sec:conclusion}

This study presents VerifAI, the first open-source expert system for biomedical question answering that combines retrieval-augmented generation with a domain-adapted verification engine into a fully integrated and modular pipeline. By grounding answer generation in hybrid document retrieval and layering it with automated claim verification, VerifAI provides accurate, reference-backed answers along with an assessment of their factual consistency.

Each component was independently fine-tuned and rigorously evaluated on domain-relevant benchmarks, and the system achieved strong end-to-end results across 178 biomedical questions. Our evaluation shows that optimizing retrieval, generation, and verification together --- rather than in isolation --- yields more robust and trustworthy outputs. The verification engine further demonstrated domain-specific advantages over large general-purpose models such as GPT-4, especially in handling nuanced biomedical inference tasks.

\hlchange{Our results demonstrate that task-specific fine-tuning allows SLMs to not merely approximate but potentially surpass generalized frontier models in structural adherence tasks like citation formatting. By constraining the model's generation to a specific referencing schema, we achieve higher citation fidelity with a fraction of the parameter count.}

Beyond this biomedical use case, VerifAI is designed as a generalizable enterprise-level framework. Its modular codebase enables straightforward adaptation to other high-stakes domains, including legal, financial, and regulatory applications --- where evidence traceability and factual correctness are essential. Provided access to a stable retrieval corpus and verification labels, users can deploy VerifAI in their own specialized contexts.

While promising, the results reveal future challenges: improving contradiction detection, expanding evaluation with domain-expert feedback, and strengthening performance on list-type or multi-fact questions. Nevertheless, VerifAI lays the groundwork for advancing trustworthy, transparent, and verifiable question answering systems in scientific and decision-critical environments.

\section{Limitations}
\label{sec:limitations}

While VerifAI demonstrates strong performance across retrieval, generation, and verification, several limitations remain. \hlchange{These challenges highlight critical areas for future investigation in the field of verifiable expert systems.}

First, our evaluation partially relies on GPT-4 Turbo as an automatic judge for answer quality and verification labels. Although this provides scalable and consistent annotation, it may introduce biases or overlook domain-specific nuances that human experts could better assess. \hlchange{Future research should focus on developing standardized, expert-annotated benchmarks for biomedical hallucination detection to calibrate and validate these automated judges more rigorously.}

Second, while our hybrid IR system improves recall, it does not guarantee exhaustive coverage of all relevant evidence. This is particularly evident for list-type questions, where missing even one relevant abstract can negatively impact both the generation and verification outcomes. \hlchange{A promising direction to address this is the development of agentic retrieval workflows that utilize iterative query decomposition, allowing systems to actively refine searches until information coverage goals are met.}

Third, the verification model still struggles to identify contradictions, especially in complex biomedical contexts where conflicting evidence may be subtle or require domain knowledge beyond the abstract level. \hlchange{Addressing this requires advancing verification architectures beyond simple entailment, potentially through chain-of-thought reasoning mechanisms or by training on datasets enriched with "hard negatives" that specifically target numerical discrepancies and implicit semantic inversions.}

Fourth, the lack of human-annotated NLI labels in biomedical benchmarks required us to generate reference labels using GPT-4, which --- although validated by standalone comparisons --- remains a proxy for true expert annotation.

Lastly, although we designed VerifAI to be modular and generalizable, our current evaluation is limited to biomedical data (PubMed). Testing in additional domains and with real-world users will be necessary to validate broader applicability.

Addressing these limitations \hlchange{through these proposed avenues---such as agentic retrieval and reasoning-enhanced verification---}will be essential for advancing VerifAI’s robustness, interpretability, and practical deployment in high-stakes environments.

\section{Code and Data Availability}
\label{sec:availability}

To promote transparency and reproducibility, all components of VerifAI are released as open-source software under the AGPL-3.0 license. The following resources are publicly available:

\textbf{Code Repository:} The full system implementation, including the Information Retrieval (IR), Generative Component (GC), and Verification Component (VC) modules, is available at \url{https://github.com/nikolamilosevic86/verifAI}. The repository includes installation scripts, configuration templates, indexing pipelines, and deployment instructions for both the biomedical (PubMed) and general-purpose (document-based) versions of VerifAI.

\textbf{Fine-tuned Models:} The QLoRA-adapted Mistral-7B-Instruct-v0.2 model for citation-aware generation is available at \url{https://huggingface.co/BojanaBas/Mistral-7B-Instruct-v0.2-pqa-10}. The DeBERTa-based verification model, fine-tuned on the transformed SciFact dataset, is available at \url{https://huggingface.co/MilosKosRad/TextualEntailment_DeBERTa_preprocessedSciFACT}.

\textbf{Training Datasets:} The PQAref dataset used for fine-tuning the generative component is available at \url{https://huggingface.co/datasets/BojanaBas/PQAref}. The transformed SciFact dataset used for training the verification component is available at \url{https://huggingface.co/datasets/MilosKosRad/SciFact_VerifAI}.

\textbf{Live Demonstration:} A fully deployed instance of VerifAI is accessible at \url{https://verifai.institutonline.ai/}, allowing users to interact with the system without local installation. Additional project information and documentation are available at \url{https://verifai-project.com/}.

All evaluation scripts, configuration files, and deployment guides are documented in the repository to facilitate replication of our results.

\bibliographystyle{elsarticle-num} 
\bibliography{cas-refs}

\begin{IEEEbiography}[{\includegraphics[width=1in,height=1.25in,clip,keepaspectratio]{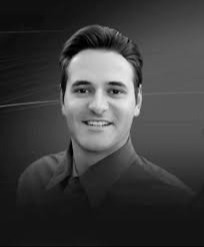}}]{Milo\v{s} Ko\v{s}prdi\'{c}} was born in Novi Sad, Serbia. He received the B.A. degree in Serbian language and general linguistics in 2012, the M.A. degree in linguistics in 2013, and the Ph.D. degree in linguistics and Serbian language in 2014, all from the Faculty of Philosophy, University of Novi Sad, Serbia. He also received a B.S. degree in mathematics from the Faculty of Sciences, University of Novi Sad, Serbia, in 2013, and an M.S. degree in computing and social sciences (2020) and a Ph.D. degree in artificial intelligence (2020) from the University of Belgrade, Serbia.

He is currently a Researcher at the Institute for Artificial Intelligence of Serbia, Novi Sad, where he works in the field of natural language processing (NLP) and large language models (LLMs), with a focus on semantic text search and scientific question answering. He is also a Senior Associate at the Linguistics Department of Petnica Science Center, Valjevo, Serbia, having previously served as Head of the department. He has published on topics including claim verification using deep learning models, natural language inference, and sentiment analysis of Serbian language texts.
\end{IEEEbiography}

\begin{IEEEbiography}[{\includegraphics[width=1in,height=1.25in,clip,keepaspectratio]{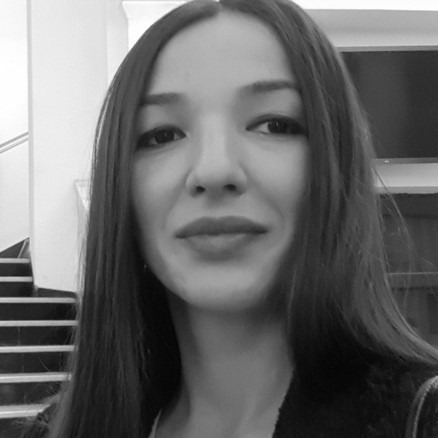}}]{Adela Ljaji\'{c}} was born in Serbia and holds an M.Sc. in Computer Science from the University of Belgrade (2007) and a Ph.D. in Computer Software Engineering from the University of Niš (2019).

She is a Research Associate at the Institute for Artificial Intelligence, Research and Development of Serbia. Her work focuses on natural language processing, large language models, and question answering, with prior experience at the Allen Institute for AI and the State University of Novi Pazar.

Her research covers LLM fine-tuning, information retrieval, retrieval-augmented generation, claim verification, named entity recognition, sentiment analysis, and topic modeling, with broader interests in semantic similarity and semantic search.
\end{IEEEbiography}

\begin{IEEEbiography}[{\includegraphics[width=1in,height=1.25in,clip,keepaspectratio]{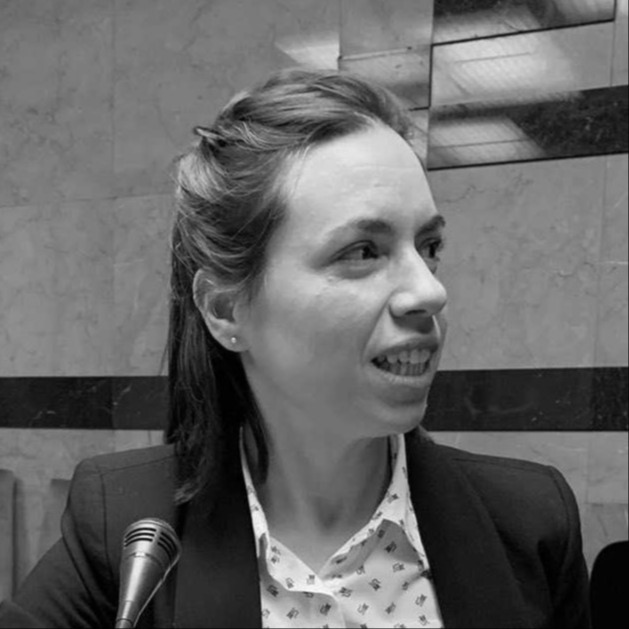}}]{Bojana Ba\v{s}aragin} was born in Serbia. She
received her Bachelor’s degree in General Linguistics in 2006 and her Doctor of Philosophy (Ph.D.) in Computational Linguistics from the Faculty of Philology, University of Belgrade, in 2017. Her doctoral research focused on the development of formal grammars for the Serbian language, including the foundations of FBLTAG for Serbian and the use of metagrammars.

She is currently a Senior Researcher at the Institute for Artificial Intelligence Research and Development of Serbia, working in the Human-Computer Interface group. Her work emphasizes natural language processing (NLP) applied to Serbian language data, particularly the creation and maintenance of linguistic resources and tools using large language models (LLMs). These include anonymizers for the Serbian language, as well as pipelines for early disease detection. She has been involved in both academic and industrial projects, ranging from healthcare-related NLP to the development of classification pipelines for the improvement of customer service in companies. Her broader research interests include sentiment analysis, topic modeling, named entity recognition, question answering, and the integration of AI methods into interdisciplinary contexts.
\end{IEEEbiography}

\begin{IEEEbiography}[{\includegraphics[width=1in,height=1.25in,clip,keepaspectratio]{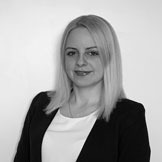}}]{Darija Medvecki} was born in Serbia in 1990. She received her  Bachelor’s and Master’s degrees from the Faculty of Technical Sciences, University of Novi Sad. She is currently pursuing the Ph.D. degree in artificial intelligence at the same faculty and serves as a Research Assistant at the Institute for Artificial Intelligence Research and Development of Serbia. Her main research interest is natural language processing (NLP) and its applications in diverse fields, including the biomedical domain, customer service, and social media. Within NLP, her current research focuses on large language models, sentiment analysis, topic modeling, named entity recognition, and the development of Serbian-language resources.

\end{IEEEbiography}

\begin{IEEEbiography}[{\includegraphics[width=1in,height=1.25in,clip,keepaspectratio]{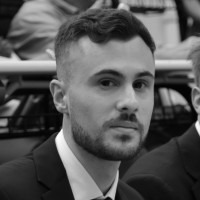}}]{Lorenzo Cassano} graduated with honors with a Bachelor's degree in Computer Science from the University of Bari Aldo Moro and a Master's degree in Artificial Intelligence from Università di Bologna. He is an enthusiastic computer scientist specializing in artificial intelligence with a strong passion for machine learning and research-driven innovation.

During his internship at Bayer Pharmaceuticals in Berlin, Lorenzo contributed to the development of the VerifAI system described in the current paper.
\end{IEEEbiography}

\begin{IEEEbiography}[{\includegraphics[width=1in,height=1.25in,clip,keepaspectratio]{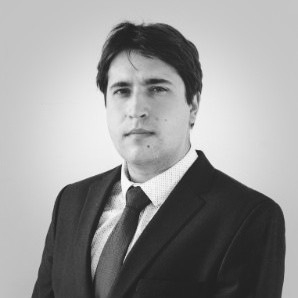}}]{Nikola Milo\v{s}evi\'{c}} was born in 1986 in Bratislava, Slovakia, and later lived in Belgrade, Serbia. He completed his undergraduate and master's studies at the University of Belgrade, Faculty of Electrical Engineering. He earned his Ph.D. in Computer Science from the University of Manchester, UK, where his research focused on text mining and natural language processing in biomedicine.

He is currently a Data Science Manager and Science Fellow at Bayer Pharmaceuticals in Berlin, leading the development of scalable NLP platforms, knowledge graphs, information retrieval, and generative AI systems. He is also a Research Fellow at the Institute for Artificial Intelligence of Serbia. Nikola is deeply involved in developing GenAI workflows, agentic systems, knowledge graphs, and information retrieval engines. His work includes applying machine learning to biomedical text mining and combating hallucinations in large language models.
\end{IEEEbiography}

\end{document}